%% file: main.tex
\documentclass[%
preprint,
nofootinbib,
 amsmath,amssymb,
 aps,
]{revtex4-1}

\usepackage{dcolumn}

\allowdisplaybreaks

\usepackage[dvipdfmx]{graphicx}
\usepackage{latexsym}
\usepackage{amsfonts}
\usepackage{amssymb}
\usepackage{amsmath}
\usepackage{bm}
\usepackage{mathrsfs} 

\newcommand{\bx}{{\bm{x}}}

\newcommand{\bq}{{\bm{q}}}

\newcommand{\Br}{B_{\rm r}}

\newcommand{\rp}{{r_\parallel^2}}
\newcommand{\rt}{{r_\perp^2}}

\newcommand{\ome}{\tilde \omega}
\newcommand{\epslim}{ \epsilon_{ {}_{\rm lim} } }
\newcommand{\nlim}{ n_{ {}_{\rm lim} } }

\newcommand{\eperp}{\epsilon_\perp}
\newcommand{\epara}{\epsilon_\parallel}

\def\simge{\mathrel{%
   \rlap{\raise 0.511ex \hbox{$>$}}{\lower 0.511ex \hbox{$\sim$}}}}
\def\simle{\mathrel{
   \rlap{\raise 0.511ex \hbox{$<$}}{\lower 0.511ex \hbox{$\sim$}}}}
\def\bigs{\mathrel{
   \rlap{\raise 0.531ex \hbox{$>$}}{\lower 0.531ex \hbox{$<$}}}}

\usepackage[normalem]{ulem}  
\usepackage[dvips]{color} 

\renewcommand\sout{\bgroup \color{red} \ULdepth=-.5ex \ULset}

\usepackage{slashed}

\begin{document} 

\input{head}


\section{Introduction}
\label{sec:intro}

\begin{figure}[b]
	\begin{center}
		\includegraphics[width=0.9\hsize]{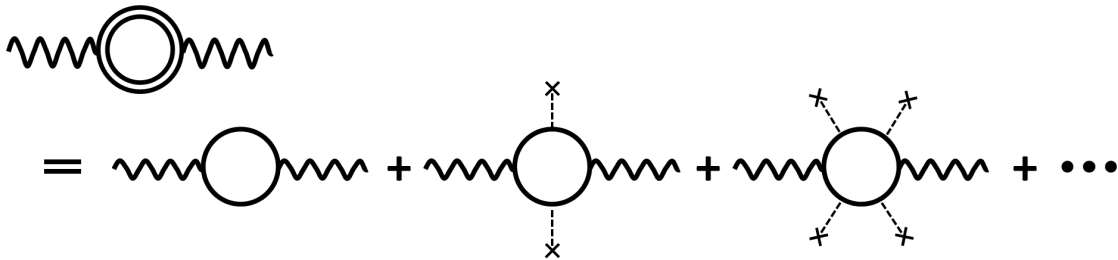}
	\end{center}
\vspace{-0.8cm}
\caption{Vacuum polarization in a magnetic field. }
\label{Fig:vp}
\end{figure}

In our recent paper  which we call ``paper I" \cite{HI1}, we analytically 
computed the vacuum polarization tensor of a propagating photon in a 
strong magnetic field at the one-loop level of a `dressed' fermion, 
which is diagramatically shown in Fig.~\ref{Fig:vp}. While the interaction of the 
fermion with the
propagating photon (wavy lines in Fig.~\ref{Fig:vp}) is considered at the 
lowest order with respect to the coupling constant, the 
resultant diagram contains 
all-order tree-level interactions with the external field (dashed lines) 
through the use of the dressed fermion (a double line). Such diagrams are 
enhanced when the external magnetic field is strong enough, because each 
insertion of the external field $B$ yields an enhancement factor 
$eB/m^2=B/B_c$ where $m$ is a mass of the fermion and 
$B_c\equiv m^2/e$ is the corresponding critical magnetic field. 
The vacuum polarization tensor, or equivalently, the self-energy of a photon 
is a fundamental quantity that carries information on the change of 
properties of a propagating photon in response to the external field. 

Having obtained the analytic representation of the polarization tensor, 
we can now investigate {\it vacuum birefringence} and {\it decay of 
a photon into a fermion-antifermion pair}, both of which 
are characteristic phenomena of ``nonlinear QED effects" and should 
become visible in external fields stronger than the critical magnetic 
field $B_c$. 
In the presence of external magnetic fields, the vacuum 
polarization tensor allows for additional terms which are absent 
in the ordinary vacuum (see Eq.~(\ref{eq:Pex})). 
These additional terms depend on polarizations of 
the photon, and consequently lead to birefringence phenomena. 
Furthermore, the polarization tensor obtained in 
paper I \cite{HI1} is expressed as infinite summation over 
all the possible pairs of Landau levels 
corresponding to those of a fermion and of an antifermion in the magnetic 
field, and show quite nontrivial behavior when the photon momentum is 
around the thresholds of the decay 
determined by the Landau levels. In particular, 
the vacuum polarization tensor has an imaginary part when the photon 
momentum is larger than the lowest threshold which indicates the 
decay of a photon into a pair of a fermion and an antifermion, 
both of which are in the lowest Landau 
levels with vanishing momenta along the direction of the magnetic field. 
Moreover, each term of the infinite summation of the polarization
tensor corresponds 
to a pair of Landau level indices, and thus has a unique threshold
determined by the Landau levels. 
In all the terms, an imaginary part appears 
when the photon momentum exceeds the threshold specific to individual terms. 

\begin{figure}[t]
	\begin{center}\hspace*{-5mm}
		\includegraphics[scale=0.65]{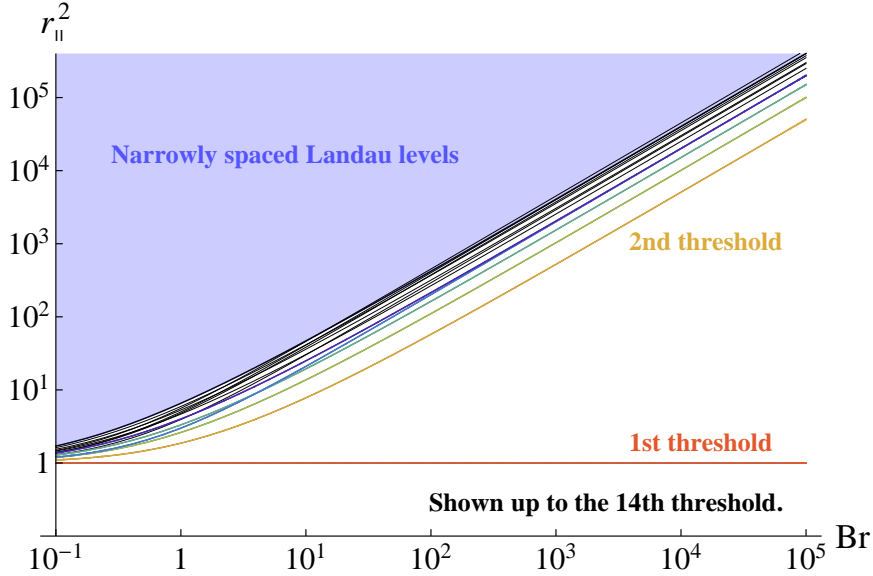}
	\end{center}
\caption{
Threshold structure of a photon momentum $r_\parallel=q_\parallel^2/(4m^2)$ 
as a function of $\Br=B/B_c=eB/m^2$. 
Curves are positions of thresholds 
$q_\parallel^2=\{\sqrt{m^2+2\ell eB}+\sqrt{m^2+2(\ell +n)eB}\}^2$ 
with $\ell,n\in {\mathbb Z}$, determined by the Landau levels with vanishing 
longitudinal momenta. The shaded region is densely 
filled with the threshold lines. The horizontal line at $\rp=1$ corresponds
to the lowest threshold ($n=\ell=0$). 
}
\label{Fig:map}
\end{figure}

In the present paper, we compute {\it refractive indices} and 
{\it dielectric constants} by using the analytic expression for 
the vacuum polarization tensor. We investigate very carefully 
their behavior 
around the thresholds, and how the imaginary part 
appears when the photon momentum exceeds the first threshold given by 
the lowest Landau levels (LLLs). In fact, since all the thresholds have 
similar structures 
as that of the first threshold in the vacuum polarization tensor, 
we focus in the present paper 
on the first threshold. In other words, we consider the results in the 
LLL approximation. This physically corresponds 
to working in 
the strong limit of the external magnetic field because the first 
threshold is independent of the magnitude of the magnetic field and thus 
gets isolated more distantly from the higher thresholds 
as the magnetic field becomes stronger. This is shown in Fig.~\ref{Fig:map}. 
Thus, our calculation in the present paper is valid
below the second lowest threshold 
$q_\parallel^2 <\left(m+\sqrt{m^2+2eB}\right)^2$.

The refractive indices $n$ and the dielectric constants $\epsilon$ 
are defined through the dispersion relations for the propagating photons
$\epsilon=n^2=|\bq^2 |/\omega^2$, 
which one obtains by solving the Maxwell equation with the vacuum 
polarization tensor included \cite{MS76,HI1}. 
Then, one finds that they are in general 
expressed by  
the scalar coefficients $\chi_i$ ($i=0,1,2$) of the vacuum polarization 
tensor (only $\chi_0$ remains nonzero in the vanishing field limit):
\begin{eqnarray}
&&\epsilon_\perp = \frac{1+\chi_0}{1+\chi_0+\chi_2 \sin^2 \theta }\, , 
\label{epsilon_perp} \\
&&\epsilon_\parallel = \frac{1+\chi_0+\chi_1}{1+\chi_0+\chi_1 \cos^2 \theta } 
\, , \label{epsilon_parallel}
\end{eqnarray}
where $\theta$ is the angle between the photon momentum and the magnetic 
field, and the perpendicular $\perp$ and parallel $\parallel$ indices imply 
polarizations with respect to the direction of the magnetic field. 
However, these expressions do not immediately provide the 
results we want. This is because the scalar functions $\chi_i$ are 
in general functions of the photon momentum $q^\mu$ and thus contain 
the refractive indices (or the dielectric constants) through the 
dispersion relation. Thus, to obtain the correct values for the 
refractive indices (and dielectric constants), 
one has to treat the equations self-consistently. We will see that 
such a self-consistent procedure gives a large effect when the deviation
of the refractive indices (or the dielectric constants) from unity is large.

As is well explained in standard textbooks on optics, refractive 
indices have clear physical meaning: 
Real part of the refraction index provides 
a {\it phase velocity} $v_{\rm phase}=1/n_{\rm real}$ of a propagating photon 
and imaginary part is called an {\it extinction coefficient} $\kappa = n_{\rm imag}$ 
for the photon propagation. 
They appear in a phase factor and a damping factor 
of the photon field $\Psi(t,\bx)$ (see Appendix~D in Ref.~\cite{HI1} for detail):
\begin{eqnarray}
\Psi (t,\bx) \propto 
\exp \{  - i \omega ( t - v_{\rm phase}^{-1}\, \hat \bq \cdot \bx ) \} \ 
\exp\{ -  \omega \, \kappa \, \hat \bq \cdot \bx \}
\ \ ,
\label{eq:photon_field}
\end{eqnarray}
where $\hat \bq$ is a unit vector directed to the photon propagation. 
Thus it is natural to define 
{\it decay length} in which the intensity of the photon field 
$I(t,\bx) \propto \vert \Psi (t,\bx) \vert^2$ falls off in a strong magnetic 
field, 
\begin{eqnarray}
d \equiv \frac{1}{2  \omega \kappa}=\frac{1}{ 2  \omega  n_{\rm imag} }
\label{eq:decay_length}
\ \ .
\end{eqnarray}
When the refractive index has an imaginary part, 
the decay length takes a finite value, 
which also indicates that the photon decays within a finite lifetime. 
Notice that these quantities depend on the photon energy $\omega$ and the 
angle $\theta$ between the magnetic field and the photon propagation direction.
In fact, they are phenomenologically important when we apply our results to 
realistic situations such as in heavy-ion collisions and high-intensity 
laser where strong magnetic fields are realized only in a small space-time 
volume. 
Namely, if decay length and lifetime 
are small enough compared to the 
spatial extent and time duration of strong magnetic fields, 
then the decay process
will be dominant. Otherwise only the birefringence phenomena will be seen.

The present paper is organized as follows: in the next section, we 
briefly review the main results obtained in paper I. 
In particular, we give the explicit form of the vacuum polarization 
tensor in the LLL approximation which is used in the present paper. 
Since the perpendicular polarization
 of the dielectric constant is not modified ($\epsilon_\perp=1$) 
in the LLL approximation\footnote{
The other contributions from the higher Landau levels barely modify 
the perpendicular polarization $\epsilon_\perp$ 
in the kinematical region below the lowest threshold. 
This will be briefly mentioned again in associate with Fig.~\ref{fig:Br}. 
}, 
we will discuss only the parallel 
polarization $\epsilon_\parallel$ in the following sections.
In Sec.~\ref{sec:generalities}, we discuss  general aspects of 
the dielectric constant in the LLL approximation, which is followed by
detailed description in Sec.~\ref{sec:self_consistent}. 
There, we compare results from different methods in solving the 
self-consistent equation for the dielectric constant. Then, 
in Sec.~\ref{sec:refractive_index},
we show the refractive index of the parallel component. Since the refractive
index is trivially related to the dielectric constant, the results are
qualitatively the same, but still, it will be useful when we apply our 
results to phenomenology. 
The last section is devoted to summary and discussions.
We point out similarities between the vacuum birefringence
in a strong magnetic field and the optical properties of substances. 
We also comment on 
future applications in laser physics, heavy-ion physics, 
and compact stars with high magnetic fields. 
In Appendices, we provide some supplementary information such as the 
dependence of dielectric constant 
on the propagation angle and the magnetic field strength, 
and a brief comment on other higher Landau levels.


\section{Vacuum polarization tensor in strong magnetic fields}
\label{sec:review}



In this section, we briefly summarize the main results obtained in paper I 
\cite{HI1}. We also provide 
notations necessary for the calculation in the present paper. 
As we mentioned in Introduction, we work in the strong field limit and thus
we present here the explicit analytic expression of the vacuum polarization 
tensor in the LLL approximation.

\subsection{Vacuum polarization tensor and dielectric constants}

Suppose that the external magnetic field is applied along the third axis 
of spatial coordinates in the negative direction so that $eB^3=|e|B>0$ for 
an electron 
($B^3=-B$ is the third component of the magnetic field vector $B^i$, 
and $e=-|e|$ is negative for an electron). 
The photon momentum $q^\mu$ is now decomposed into longitudinal and 
transverse components with respect to the magnetic field, namely 
$q_\parallel^{\mu} = (q^0, 0, 0, q^3)$ and 
$q_\perp^{\mu} = (0, q^1, q^2, 0)$. Correspondingly, 
the metric tensor $\eta^{\mu\nu} = {\rm diag} (1,-1,-1,-1)$ is also 
decomposed into longitudinal and transverse subspaces 
$\eta_\parallel^{\mu\nu} = {\rm diag} (1,0,0,-1)$ and 
$\eta_\perp^{\mu\nu} = {\rm diag} (0,-1,-1,0)$. 

In the absence of the full Lorentz symmetry, the vacuum polarization tensor 
$\Pi_{\rm ex}^{\mu\nu} (q_\parallel, q_\perp; \Br)$ allows for two 
additional terms. Using the projection 
operators $P_i^{\mu\nu}$ satisfying $q_\mu P_i^{\mu\nu}=0$, 
a gauge-invariant tensor structure is given by 
\cite{Adl1, Tsa, TE, MS76, Urr78} (see also Ref.~\cite{HI1} and Refs.~\cite{DR,DG} for details), 
\begin{eqnarray}
&&
\Pi_{\rm ex}^{\mu\nu} (q_\parallel, q_\perp; \Br) = - \Big( \chi_0  P_0^{\mu\nu} + \chi_1 P_1^{\mu\nu} + \chi_2  P_2^{\mu\nu} \Big)\, ,
\label{eq:Pex}
\\
&&
P^{\mu\nu}_0 = q^2 \eta^{\mu\nu}  - q^\mu q^\nu 
\ , \ \ 
P^{\mu\nu}_1 = q^2_\parallel  \eta^{\mu\nu}_\parallel  - q^\mu_\parallel  q^\nu_ \parallel 
\ , \ \ 
P^{\mu\nu}_2 = q^2_\perp \eta^{\mu\nu}_\perp - q^\mu_\perp q^\nu_\perp
\label{eq:Pmn}
\, ,
\end{eqnarray}
where we suppressed arguments of Lorentz-scalar coefficient functions 
$\chi_i \ (i=0,1,2)$. 
All the three coefficient functions depend on the magnetic field $\Br=B/B_c$ 
and 
the photon momentum, $q^\mu=q^\mu_{\parallel}+q^\mu_{\perp}$. 
One can show that the vacuum polarization tensor in the ordinary vacuum is appropriately reproduced 
in the vanishing magnetic field limit: 
$\chi_0$ results in the coefficient function of the vacuum polarization tensor in the ordinary vacuum, 
while the other two $\chi_{1,2}$ vanish. 

Plugging the general form of the polarization tensor (\ref{eq:Pex}) into the 
Maxwell equation for the propagating photon, we can solve it to obtain the 
dispersion relations for physical modes (see Sec.~3 and Appendices~C and D 
in paper I \cite{HI1}). As mentioned in Introduction, 
the dielectric constants $\epsilon$ or 
the refractive indices $n$ are defined by 
\begin{equation}
\epsilon  = n^2= \frac{ | \bq |^2 }{  \omega^2}\, ,
\label{eq:eps_def}
\end{equation} 
where the photon momentum is $q^\mu=(\omega, \bq)$. 
Then, one finds two different values $\epsilon_\perp$ and $\epsilon_\parallel$ 
for the dielectric constants as shown in Eqs.~(\ref{epsilon_perp}) and 
(\ref{epsilon_parallel}).
There, we have taken a coordinate system in which a photon is propagating in 
the $y=0$ plane so that the momentum vector $q^\mu$ is represented as 
$q^\mu = ( \omega, |\bq| \sin(\pi-\theta), 0, |\bq| \cos (\pi -\theta)  )$ 
with $\theta$ being the angle between the direction of the external magnetic 
field and the momentum of a propagating photon.

Since two dielectric constants $\epsilon_\perp$ and $\epsilon_\parallel$,
and consequently, two refractive indices $n_\perp$ and $n_\parallel$, are 
in general not equal to unity and different from each other, 
we call this phenomenon {\it vacuum birefringence} after a similar 
phenomenon in dielectric substances. 
Notice also that, due to the violation of the Lorentz symmetry by an 
external magnetic field, the dielectric constants explicitly depend on 
the zenith angle $\theta$. Nevertheless, the system maintains boost 
invariance in the third direction, and thus photon propagations in 
the directions at $\theta =0, \pi$ are special. Indeed, substituting these 
angles into Eqs.~(\ref{epsilon_perp}) and (\ref{epsilon_parallel}), 
we find that both of them become unity 
$\epsilon_\perp(\theta=0,\pi)=\epsilon_\parallel(\theta=0,\pi) =1$ and that 
there is no effect from the magnetic field. 

As we will discuss later and also discussed in detail in paper I \cite{HI1}, 
the scalar coefficients can become complex when the photon momentum exceeds 
some threshold values. In that case, the corresponding dielectric constants 
and 
refractive indices are also complex. Since the scalar coefficients $\chi_i$ 
are parts of the self-energy of the photon, emergence of an imaginary part 
indicates the decay of a photon into a fermion-antifermion pair.


\subsection{Scalar coefficient functions in the LLL approximation}

Let us outline how to obtain analytic representations for 
the scalar coefficient functions $\chi_i$ in the external magnetic fields. 
In paper I,  we computed $\chi_i$ by 
Schwinger's proper time 
method which is a very useful technique for the calculations under 
external fields \cite{Sch}. 
Each coefficient function was first expressed as a double 
integral with respect to two proper times, each of which is associated 
with the fermion propagator composing the fermion loop 
(see Fig.~\ref{Fig:vp}). Then, the integration was analytically 
performed after we made rapidly oscillating
integrands tractable by expanding them with known special functions.
The results we obtained are represented as infinite summation with 
respect to two indices, which are afterward identified with the 
Landau levels. In particular, the first term in the double infinite 
sum corresponds to the contribution of the lowest Landau levels, 
which was confirmed to agree with expressions obtained by 
other approximate methods in the strong field limit \cite{MS76,Fuk}. 
We will investigate this contribution in detail in the present work. 
See paper I for details of the 
calculation and the explicit analytic forms of all the scalar coefficient 
functions \cite{HI1}. 

Here we just show the explicit form of the coefficient functions from 
the lowest Landau levels. Below, we use the following notations for 
dimensionless 
kinematical variables: $r_\parallel^2 = { q_\parallel^2 }/{ (4m^2) }$ and 
$r_\perp^2 = { q_\perp^2 }/{(4m^2) } = - { | \bq_\perp | ^2 }/{ (4m^2) }$. We also use a shorthand notation: $\eta \equiv - 2 { r_\perp^2 }/{ \Br } $. 
By taking the first term in the sum with respect to the Landau levels 
(see Sec.~4.2 in paper I~\cite{HI1} for details), 
we find that the coefficient functions $\chi_i$ 
of the vacuum polarization tensor survive only for $i=1$:
\begin{eqnarray}
&&\chi_0^{\rm LLL} = \chi_2^{\rm LLL} = 0
\label{eq:chi02_LLL}
\, ,\\
&&\chi_1^{\rm LLL}(\rp, \rt ; \Br) 
= \frac{\alpha \Br }{4\pi } \ {\rm e}^{-\eta}  
\times \frac{1}{r_\parallel^2} \left\{ I_{0 \Delta}^0 (r_\parallel^2) - 2  
\right\}
\label{eq:chi1_LLL}
\, ,
\end{eqnarray}
where $\alpha=e^2/(4\pi)$ and $ I_{0 \Delta}^0 (r_\parallel^2)$ is a function
responsible for the threshold behavior at the lowest Landau level. 
Note that the dynamics encoded in the above expression is interpreted 
in terms of the factorized dependence on the photon momenta, $\rp$ and $\rt$. 
A Gaussian factor, $ \frac{\vert eB \vert }{2 \pi} 
\cdot \exp( \, - \vert{\bm q_\perp}\vert^2/(2\vert eB \vert) \, ) $, 
corresponds to the wave function of an electron in the lowest Landau level 
representing the degeneracy factor for the angular momentum and the transverse extension, 
while the rest part agrees with the free photon vacuum polarization tensor 
in 1+1 dimensions. This structure reflects the motion of an electron in 
strong magnetic fields restricted in the longitudinal direction, which is 
termed ``dimensional reduction" 
(see, e.g., Refs.~\cite{Fuk,GMS96} and references therein).

\begin{figure}[t]
  \begin{center}
   \includegraphics[width=0.6\hsize]{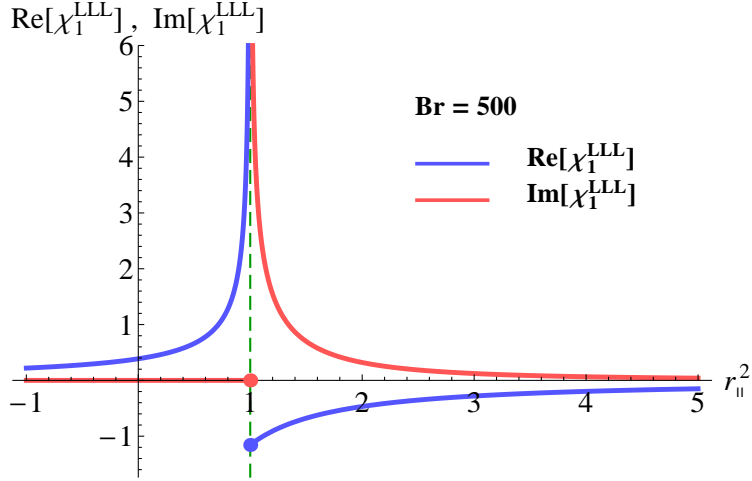}
  \end{center}
\vspace{-5mm}
\caption{
The scalar function $\chi_1^{\rm LLL}$: 
blue and red lines show real and imaginary parts of $\chi_1^{\rm LLL}$. 
}
  \label{fig:chi1_LLL}
\end{figure}

Depending on the value of $r_\parallel^2$, 
the piecewise expression of $I_{0 \Delta}^0 (r_\parallel^2)$ is given as 
\begin{eqnarray}
I_{0 \Delta}^0 (r_\parallel^2)
= 
\left\{
\begin{array}{l}
\frac{1}{ \sqrt{ r_\parallel^2(r_\parallel^2-1) } }
\ln \left| 
\frac{ r_\parallel^2 - \sqrt{  r_\parallel^2(r_\parallel^2-1)  } }{ r_\parallel^2 + \sqrt{  r_\parallel^2(r_\parallel^2-1)  }} 
\right|
\ \hspace{24mm} ( r_\parallel^2  < 0  )
\\ 
\frac{2}{\sqrt{ r_\parallel^2( 1 - r_\parallel^2) } }
 \arctan \left\{ \frac{r_\parallel^2}{\sqrt{r_\parallel^2(1-r_\parallel^2)} } \right\}
\ \qquad\ \qquad( 0< r_\parallel^2 < 1 )
\\ 
\frac{1}{ \sqrt{ r_\parallel^2(r_\parallel^2-1) } }
\left[ \ 
\ln \left| 
\frac{ r_\parallel^2  - \sqrt{r_\parallel^2(r_\parallel^2-1) } }{ r_\parallel^2  + \sqrt{r_\parallel^2(r_\parallel^2-1) }} 
\right|
+  \pi i
\ \right]
\ \qquad ( 1 < r_\parallel^2  )
\end{array}
\right.
\label{eq:I_LLL}
\, .
\end{eqnarray}
Notice that this function has a singular behavior at $\rp=1$ and that 
$\rp=1$ corresponds to the lowest threshold for the decay of a photon 
into a fermion-antifermion pair. This is confirmed by the following 
observations. 
Firstly, the function $I_{0 \Delta}^0 (r_\parallel^2)$, and thus the 
scalar function $\chi_1^{\rm LLL}$, has an imaginary part in the region, 
$\rp>1$. 
Next, as briefly commented in Introduction and shown in 
paper I \cite{HI1}, each term in the infinite summation of the 
coefficient functions is specified by two integers $n$ and $\ell$, 
and acquires an imaginary part 
as the photon momentum increases beyond the threshold value 
$\rp = \left\{\sqrt{1+2\ell \Br}+\sqrt{1+2(\ell + n)\Br}\right\}^2/4$.
This representation of the threshold suggests that 
$\ell$ and $\ell+n$ can be interpreted as the Landau levels of a fermion 
and an antifermion. From these, one finds that the condition $\rp=1$ 
is realized by the lowest Landau levels $\ell=n=0$, 
and thus emergence of an imaginary part in $\rp>1$ implies the decay 
of a photon into a fermion-antifermion pair in the lowest Landau levels. 
Note also that $I_{0 \Delta}^0 (r_\parallel^2)$ does not depend on 
$\Br$, reflecting the fact that the location of the threshold for 
the lowest Landau level, $\rp=1$, is independent of $\Br$. 
Therefore, within the LLL approximation, 
the coefficient function $\chi_1^{\rm LLL}$ linearly scales 
as the magnetic-field strength $\Br$ increases 
when $\rt \ll \Br$ (see Eq.~(\ref{eq:chi1_LLL})), 
which will be also seen 
in Fig.~\ref{fig:Br} when the photon momentum is below the lowest threshold. 
Finally, it should be emphasized that $\rp=1$ is the 
only one singular point
of the function $I_{0 \Delta}^0 (r_\parallel^2)$. Indeed, it is 
continuous at $\rp=0$ with a finite value $ I_{0 \Delta}^0 (0) = 2$, 
without any singularity. 
In Fig.~\ref{fig:chi1_LLL}, we show the behavior of the scalar function
$\chi_1^{\rm LLL}$ around the threshold $\rp=1$. The other thresholds 
from higher Landau levels have qualitatively the same structure as 
shown in Fig.~\ref{fig:chi1_LLL}. 

\section{Generalities on dielectric constant in the LLL approximation} 
\label{sec:generalities}


By using the analytic expression of the vacuum polarization tensor, 
we can compute the dielectric constants given in Eqs.~(\ref{epsilon_perp}) 
and (\ref{epsilon_parallel}). In this section, we discuss some general 
aspects of the dielectric constants and the difficulties in computing 
them from the coefficient functions $\chi_i$. A complete description 
of the dielectric constant within the LLL approximation will be 
presented in the next section.

First of all, we emphasize that 
we have not yet 
specified any dispersion relation for the external photon 
momentum in the scalar coefficient functions $\chi_i$. 
For an on-shell photon, the dispersion relations should be determined by 
Eqs.~(\ref{epsilon_perp}) and (\ref{epsilon_parallel}), 
which do contain the dielectric constants on the right-hand sides 
through the photon momenta, $\rp$ and $\rt$. According to the definition 
of the dielectric constant (\ref{eq:eps_def}), those photon momenta are 
written as, $\rp = \ome^2 (1- \epsilon \, \cos^2 \theta)$ and 
$\rt = - \epsilon \, \ome^2 \sin^2 \theta$, commonly for $\epsilon_{\perp}$ 
and $\epsilon_{\parallel}$, where we introduced a scaled photon energy, 
$\ome^2 = \omega^2/(4m^2)$. This fact indicates that we have to solve 
these relations in a self-consistent way with respect to the dielectric 
constant appearing on the both sides. As we will find later, there is 
a large deviation from the massless-type dispersion relation indicating 
that $\rp+\rt \neq 0$ and $\omega\neq |\bq |$. We will see that a 
self-consistent treatment plays an important role in such a case. 
In particular, we will investigate very carefully how the presence of 
the threshold and an imaginary part in $\chi_i$ affects the results,
which can be well studied in the LLL approximation. 

Within the LLL approximation, only $\chi_1$ is nonzero as shown 
in Eqs.~(\ref{eq:chi02_LLL}) and (\ref{eq:chi1_LLL}). 
It immediately follows from Eq.~(\ref{epsilon_perp}) 
that the dielectric constant in one of the polarization modes 
is intact, $\epsilon_\perp = 1$. This is understood from following two 
observations. First, as shown in Appendix D in paper I~\cite{HI1}, 
an electric field induced in this polarization mode oscillates 
perpendicularly to the external field. Second, since the discretized 
transverse 
momentum of a fermion is fixed to the lowest Landau level, motion of 
a vacuum pair excitation is restricted only along the external magnetic 
field. Because the direction of the possible dipole excitation is 
perpendicular to the electric field accompanying this polarization mode, 
this mode does not induce an electric dipole excitation. 
Therefore, its propagation is not modified. 
On the other hand, an electric field accompanying 
the other polarization mode has a parallel component oscillating along the 
external magnetic field. Thus, an electric dipole can be induced as a 
response of the Dirac sea, which results in a modification of the 
dielectric constant $\epsilon_\parallel$ shown in 
Eq.~(\ref{epsilon_parallel}).

Inserting $\chi_0=\chi_2=0$ into the general expressions 
(\ref{epsilon_perp}) and (\ref{epsilon_parallel}), we find 
the dielectric constants in the LLL approximation:
\begin{eqnarray}
&&
\epsilon_\perp^{\rm LLL} = 1\, ,
\label{eq:eps_LLL_t}
\\
&&
\epsilon_\parallel^{\rm LLL} (\tilde \omega, \theta; \Br) 
= \frac{ 1 + \chi_1^{\rm LLL} (\rp,\rt; \Br) }{ 1 +  \chi_1^{\rm LLL}
(\rp,\rt; \Br) \times \cos^2 \theta}
\, .
\label{eq:eps_LLL}
\end{eqnarray}
Here, we have explicitly shown the arguments of 
$\epsilon_\parallel^{\rm LLL}$ and $\chi_1^{\rm LLL}$. The dielectric constant 
$\epsilon_\parallel^{\rm LLL}$ depends on a scaled 
photon energy 
$\tilde \omega^2 \equiv \omega^2/(4m^2)$ and angle $\theta$ separately, 
because of an explicit $\theta$-dependence on the right-hand side. 
It should be noticed that photon's squared momenta 
on the right-hand side can be rewritten in terms of the scaled 
energy 
$\tilde \omega$ so that they contain the dielectric constant 
$\epsilon_\parallel^{\rm LLL}$ according to the definition (\ref{eq:eps_def}):
\begin{eqnarray}
&&\rp = \tilde \omega^2 ( 1- \epsilon_\parallel^{\rm LLL} \cos^2 \theta)\, ,
\label{rp_LLL}
\\
&&\rt = - \tilde \omega^2 \epsilon_\parallel^{\rm LLL} \sin^2 \theta\, .
\label{rt_LLL}
\end{eqnarray} 
The expression of the dielectric constant (\ref{eq:eps_LLL}) 
has to be solved with respect to $\epara$. 
We will discuss in detail that this procedure plays an important role 
when modification of the dielectric constant becomes sizable.

The vacuum birefringence in the LLL approximation bears an analogy to 
`uniaxial' birefringent material, of which optical axis corresponds to 
a preferred orientation provided by the external magnetic field. 
In this analogy, the polarization mode having the dielectric constant 
$\epsilon_\parallel$ could be called ``{\it extraordinary mode}", 
while the other one having $\epsilon_\perp$ ``{\it ordinary mode}". 
However, this analogy holds only in the LLL approximation because 
the perpendicular part $\epsilon_\perp$ in general depends on the angle 
$\theta$. Thus, if one looks at the wide range of kinematical region, 
the vacuum birefringence is much more complicated than 
the birefringence in uniaxial materials.


\subsection{Below threshold: Accuracy check of the LLL approximation}

Consider first the region below the threshold $\rp<1$. 
As long as the photon momentum is in this region, 
one can {\it numerically} perform the double integration in a safe 
way.\footnote{Each function $\chi_i$ contains a double integral 
with respect to two proper-time variables $\tau_1$ and $\tau_2$, 
and the integration over $\tau=eB(\tau_1+\tau_2)/2$ in general 
suffers from strong 
oscillation of the integrand. However, it can be shown that for $\rp<1$ 
an integral contour in the complex $\tau$-plane can be safely moved to 
the imaginary axis to obtain 
a well-behaved real-valued integral which allows for numerical 
evaluation \cite{KY,Adl1}.}
Since the original double integrals are directly evaluated 
(without being expanded with respect to special functions), 
the numerical results should contain all the contributions from 
Landau levels in our analytic expression obtained in paper I \cite{HI1}. 
Such numerical integration was performed in Ref.~\cite{KY} some years ago. 
Thus, we can check the validity of the LLL approximation by 
comparing the analytic results (\ref{eq:chi1_LLL}) \& (\ref{eq:I_LLL}) 
with the numerical results.

\begin{figure}[t]
  \begin{center}
   \includegraphics[width=0.7\hsize]{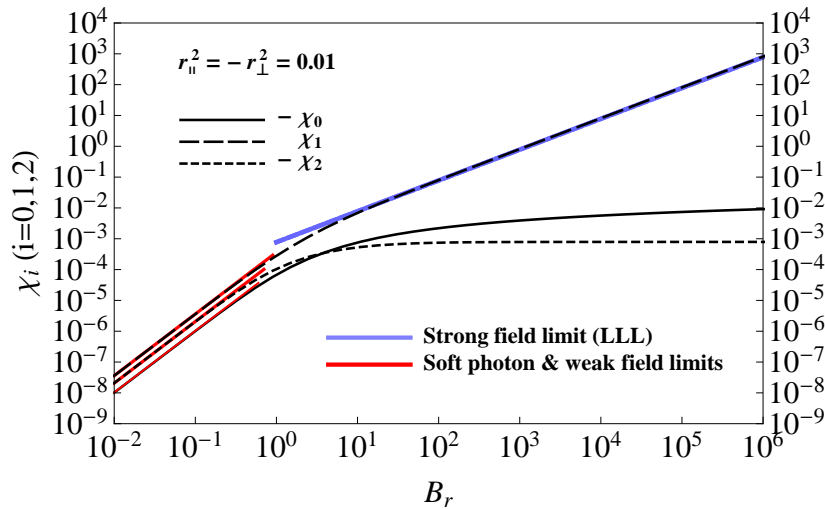}
  \end{center}
   \vspace{-8mm}
\caption{
$\Br$-dependence of the scalar functions $\chi_i \ (i=0,1,2)$: 
three black lines show results from numerical integration \cite{KY}, 
and red and blue lines show the weak-field \cite{Adl1} and strong-field 
(LLL) limits, respectively. 
}
  \label{fig:Br}
\end{figure}

In Fig.~\ref{fig:Br}, we compare the LLL results (\ref{eq:chi1_LLL}) 
\& (\ref{eq:I_LLL}) 
for the coefficient function $\chi^{\rm LLL}_1$ 
with $\chi_i$ from the numerical integration below the threshold. 
The results of numerical integration (three black lines) 
shown in Fig.~\ref{fig:Br} were reproduced by ourselves, 
and agree with the previous results in Ref.~\cite{KY} up to small 
invisible numerical errors. We also show an approximate expression 
for soft photon and weak-magnetic-field limit \cite{Adl1}. 
The photon momenta are taken to be small $\rp = - \rt = 0.01 $
below the threshold, as an illustration. 
In fact, this is in the `soft gamma-ray' regime\footnote{
Electron mass is assumed here for the fermion mass ``$m$" in $\rp$ and $\rt$. 
} 
which corresponds to typical burst spectrum 
from the soft-gamma-ray repeaters as known as magnetars. 
The focus of Ref.~\cite{KY} was the radiation from magnetars, 
which is 
the reason why they discussed the region below the threshold. 
The figure shows that 
the result of $\chi_1$ from numerical integration agrees with analytic 
expressions in the weak and strong field limits. 
In particular, the LLL approximation in $\chi_1$ works well for $\Br \agt 10$, 
and the values of $\chi_0$ and $\chi_2$ are very small compared to unity, 
both of which confirm the use of the LLL approximation 
(\ref{eq:chi02_LLL}) -- (\ref{eq:chi1_LLL}) in strong magnetic fields. 
As mentioned below Eq.~(\ref{eq:I_LLL}), we find that 
$\chi_1$ linearly scales as $\Br$ increases in the strong magnetic field 
limit.

As long as the photon momentum is below the threshold $\rp <1$, it is not 
technically difficult to obtain the dielectric constant and the 
refractive index. They have only real parts, and we can 
use the coefficient functions obtained by numerical integration. 
This was already performed in Ref.~\cite{KY}, where 
the results show that the refractive index in the parallel mode, 
$n_\parallel$,  
deviates from unity as the photon momentum approaches the threshold 
$\rp =1$ from below. 
(Essentially the same result is shown in Fig.~\ref{fig:eps0} 
for the dielectric constant in the parallel mode $\epsilon_\parallel$.)  
Note that this method is valid for arbitrary strength of the external 
magnetic field, whose validity is however 
restricted to the 
kinematical region $\rp <1$. 
It was also shown that deviation 
of $n_\parallel$ from unity 
becomes larger 
as the magnetic field strength increases beyond the critical 
field strength $\Br=B/B_c \agt 1$. 
On the other hand, the refractive index is barely modified 
in the weak field region below the the critical field strength $\Br \alt 1$, 
when the photon momentum is below the threshold $\rp < 1$. 
We will compare 
the results from the numerical integration 
with those from the LLL approximation in later sections.

The refractive index in the perpendicular mode, $n_\perp$, 
was also examined for $\rp <1$ in Ref.~\cite{KY}, 
and the result was reproduced by the present authors. 
The results do not show sizable deviation from unity, 
even when the magnetic field strength becomes larger than the critical field strength 
and when the photon momentum approaches the lowest threshold. 
Figure~\ref{fig:Br} shows that $\chi_0$ and $\chi_2$ 
are  smaller than unity by two orders in magnitude, 
and thus they bring in only tiny modification in 
the dielectric constant $\epsilon_\perp$ in Eq.~(\ref{epsilon_perp}), 
and also in the refractive index $n_\perp$. 
These results are quantitatively consistent with 
the dielectric constant (\ref{eq:eps_LLL_t}) in the LLL approximation. 

\subsection{Beyond threshold: Appearance of imaginary part}

Before we present complete self-consistent treatment of the dielectric 
constant, we explain how the dielectric constant behaves around the 
threshold in a very rough estimate. First of all, let us briefly recall
the behavior of $\chi_1^{\rm LLL}$ given in Eqs.~(\ref{eq:chi1_LLL}) and 
(\ref{eq:I_LLL}) around the threshold. 
As described below Eq.~(\ref{eq:I_LLL}), the function $\chi_1^{\rm LLL}$ has 
a singularity at the threshold $\rp=1$. When $\rp$ 
goes beyond 1, it starts to have an imaginary part. 
The real and imaginary parts of $\chi_1^{\rm LLL}$ diverge 
when the photon momentum approaches the threshold $\rp=1$ 
from below and above, respectively (see Fig.~\ref{fig:chi1_LLL}). 
Whereas the emergence of an imaginary part in $\chi_1^{\rm LLL}$ directly 
affects the dielectric constant $\epsilon_\parallel^{\rm LLL}$ (namely, 
$\epsilon_\parallel^{\rm LLL}$ starts to have an imaginary part above the 
threshold), we should notice that the dielectric constant 
takes a real finite value\footnote{This is trivially true as long as 
$\cos^2 \theta$ is not zero, $\theta\neq \pm\pi/2$. 
In two particular cases at $\theta=\pm \pi/2$, 
the dielectric constant is given by 
$\epsilon_\parallel = 1 + \chi_1^{\rm LLL}$ with $\rp = \tilde \omega^2$, 
which directly reflects the behavior of $\chi_1^{\rm LLL}$, 
and is also real below the threshold. 
These angle-dependent behaviors will be again mentioned in Sec.~\ref{subsec:angle}. 
}
at the threshold owing to cancellation of the 
divergences between the numerator and denominator in 
Eq.~(\ref{eq:eps_LLL}). In fact, we find a simple expression for 
the value of $\epsilon_\parallel$ at the threshold 
(for $\cos\theta\neq 0$): 
\begin{eqnarray}
\lim_{\rp \rightarrow 1 \pm 0 } \epsilon_\parallel (\rp) = 
\frac{ 1 }{ \cos^2 \theta }
\equiv \epslim 
\ \ ,
\label{eq:eps_limit}
\end{eqnarray}
when $\rp$ approaches the threshold from both below and above. 
Note that  the limiting value (\ref{eq:eps_limit}) 
is valid even if we include contributions from other 
levels $(\ell \neq 0, \ n \neq 0)$. This is because the divergent 
contribution to $\chi_1$ at the threshold dominates any finite 
contribution from other levels (That is why we do not put ``LLL" 
in the above equation). The same is true for the other higher
thresholds: the limiting value of $\epsilon_\parallel$ at each threshold 
comes from the contribution in $\chi_i$ which gives divergence 
at that point.

\begin{figure}
   \begin{center}
   \includegraphics[width=0.7\hsize]{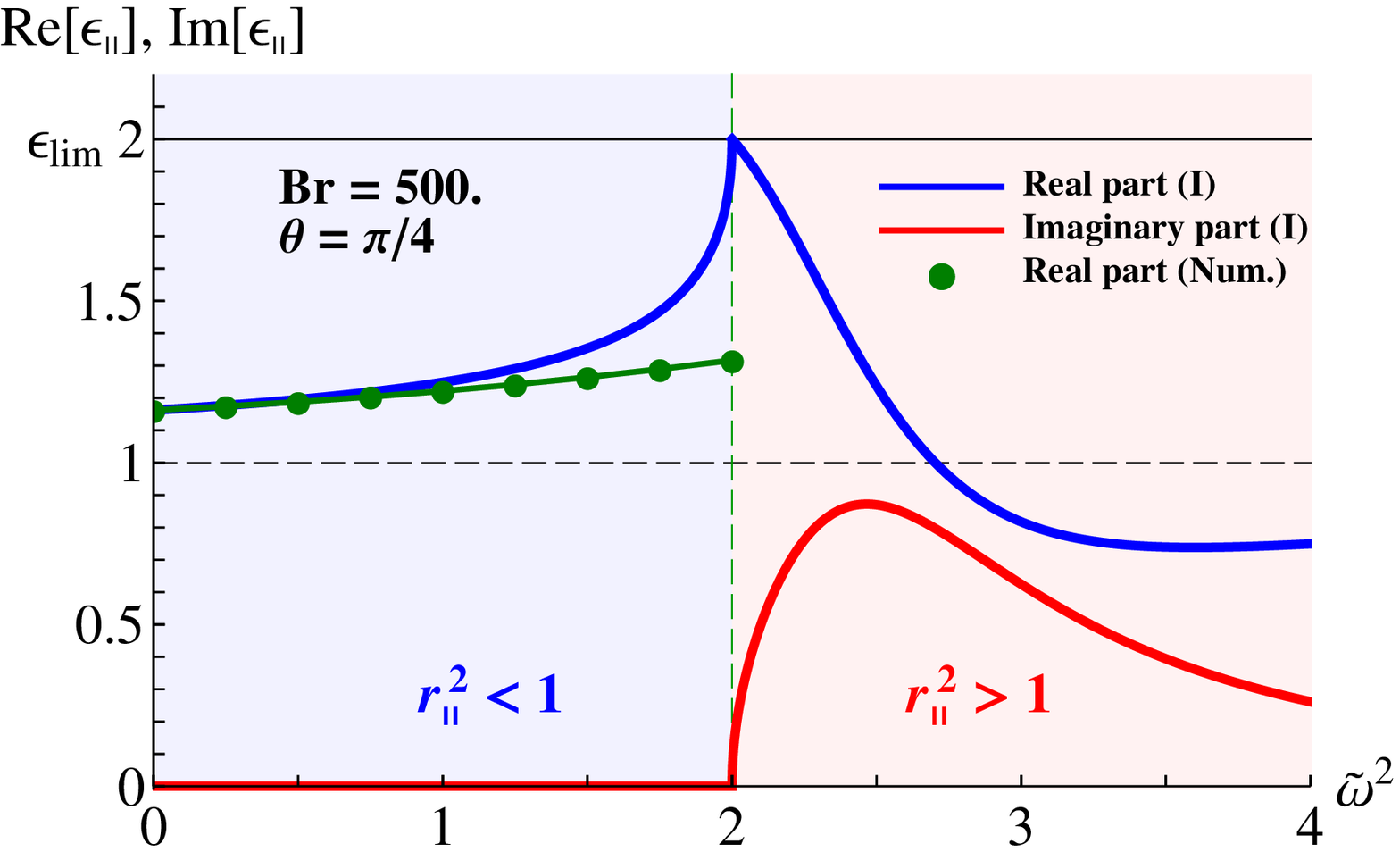}
  \end{center}
\vspace{-5mm}
\caption{
Dielectric constant in the LLL approximation: 
blue and red lines show the real and imaginary parts of the right-hand side 
of Eq.~(\ref{eq:eps_LLL}) with $\epsilon_\parallel^{\rm LLL}=1$ in 
$\chi_1^{\rm LLL}$ (called 
solution (I) in Sec.~\ref{subsec:sols}) 
Filled backgrounds indicate the kinematical regimes below and above $\rp=1$ 
whose boundary is given by $\ome^2 = 1/( 1- \cos^2 \theta )$ assuming 
$\epsilon_\parallel^{\rm LLL} = 1$ in $\rp$. 
Green dots show the numerical computation of the real part in $\rp < 1$. 
}
  \label{fig:eps0}
\end{figure}

Figure~\ref{fig:eps0} shows a plot of the dielectric constant 
(\ref{eq:eps_LLL}) obtained with $\chi_1$ shown in  
Eqs.~(\ref{eq:chi1_LLL}) and (\ref{eq:I_LLL}) as a function of the photon 
energy squared $\tilde \omega^2$. We have taken $\theta=\pi/4$ just for 
illustration (then, $\epsilon_{\rm lim}=2$), and a relatively large value for
the magnetic field $\Br=B/B_c=500$ so that 
the LLL approximation is appropriate. 
The dielectric constant which implicitly appears on 
the right-hand side of Eq.~(\ref{eq:eps_LLL}) is, at this moment, 
taken to be one. 
Then, inserting $\epsilon_\parallel^{\rm LLL} = 1$ into $\rp$ given by 
Eq.~(\ref{rp_LLL}), one finds that 
the threshold condition $\rp=1$ is now represented in the energy space 
as\footnote{Note that this equation is true only in the naive 
prescription 
where we assume $\epsilon_\parallel^{\rm LLL} = 1$ in $\rp$.} 
$\ome^2 = 1/( 1- \cos^2 \theta )=2$ 
for $\theta=\pi/4$. The region $\ome^2<2$ filled with light blue color 
in Fig.~\ref{fig:eps0} 
corresponds to the one below the threshold, while the region $\ome^2>2$
with pale red, beyond the threshold.

First let us see the region below the threshold. 
Green dots correspond to the dielectric constant $\epara$ shown in 
Eq.~(\ref{epsilon_parallel}) obtained on the basis of the 
numerical integration of $\chi_0$ and $\chi_1$ 
performed below the threshold \footnote{
To remove the divergence of $\chi_0$, 
we have employed a cut-off and renormalization condition 
described in the Appendix~B in paper I~\cite{HI1}. 
} (the same result as obtained in Ref.~\cite{KY}). 
On the other hand, a blue line corresponds to the result from 
the LLL approximation. 
As we already discussed in the previous subsection, both results 
grow with increasing photon energies. We will discuss the deviation 
between these two in the next subsection. The imaginary 
part shown with a red line is exactly zero 
in this region.

Next, look at the region beyond the threshold, $\ome^2>2$, in 
Fig.~\ref{fig:eps0}. It is clearly seen that the dielectric constant 
has a sizeable imaginary part. This contribution is never described by the 
result from numerical integration which is valid only below the threshold. 
We also find that the real part shows nontrivial behavior: 
starting from Re$[\epsilon_\parallel^{\rm LLL}]=2$ at $\ome^2=2$, it 
decreases, and even becomes 
smaller than unity. Considering the deviation between the results 
from the numerical integration and the LLL approximation in the region 
below the threshold, we expect these results (without self-consistent 
treatment) will not be quantitatively accurate. We will indeed see in 
the next section that the solution will be greatly modified in the 
self-consistent treatment. What is sure at this moment is the presence 
of the imaginary part in this region.

\subsection{Necessity of self-consistent description} 
\label{Necessity}

Let us come back to the results below the threshold. We have 
seen that there is deviation between the results from the numerical 
integrations (green dots) and the LLL approximation (a blue line), and 
that it becomes larger as the photon energy approaches the threshold. 
Since the results grow with increasing photon energies, one can 
also 
say that two results show a good agreement when the 
dielectric constant is close to unity, but deviate from each other
when it becomes large. Clearly, the analytic result with a 
large value of $\epsilon_\parallel>1$ is not consistent with 
the prescription assuming $\epsilon_\parallel=1$ on the right-hand 
side of Eq.~(\ref{eq:eps_LLL}). On the other hand, to obtain the 
result from numerical integration, Eq.~(\ref{epsilon_parallel}) was 
solved with respect to $\omega$ with its dependence on the 
right-hand side taken into account. Considering the facts that 
the result from numerical integration should be accurate in this 
region and that the magnetic field is strong enough for the 
coefficient functions to be approximated by the LLLs, we can easily
conclude that the deviation of two results nearby the threshold 
originates from the ``inconsistent" treatment of the dielectric constant
in the naive prescription.

There is another evidence for the importance of 
the self-consistent treatment. 
Note that the analytic expression of the limiting behavior 
shown in Eq.~(\ref{eq:eps_limit}) 
is obtained from the divergent behavior of $\chi_1$ when the 
longitudinal momentum approaches the threshold, $\rp \rightarrow 1$. 
Inserting the limiting 
value $\epslim$ into $\rp = \ome^2 ( 1- \epsilon_\parallel \cos^2 \theta )$, 
we however notice that, as long as $\ome$ is finite, $\rp$ approaches zero 
when the dielectric constant approaches the limiting value $\epslim$, 
\begin{eqnarray}
\lim_{\epsilon_\parallel \rightarrow \epslim} \rp( \epsilon_\parallel) 
= 0
\ \ .
\label{eq:rp_lim}
\end{eqnarray}
Therefore, Eqs.~(\ref{eq:eps_limit}) and (\ref{eq:rp_lim}) cannot be 
consistent with each other. 
This occurs when the dielectric constant is large, and must be 
resolved if we treat it self-consistently.


Then, what kind of physics is involved in the self-consistent 
description of Eq.~(\ref{eq:eps_LLL})? 
It is helpful to recall a microscopic picture of the propagation 
of a photon in an ordinary medium. Variation of the dielectric
constant or the refractive index is induced by the response of 
the medium to the incident photon. In the present case, medium is 
the vacuum which is filled with fermions in the Dirac sea, and 
the incident photon creates a polarization in it 
and, when its energy is large enough, even an on-shell fermion 
and antifermion pair which might be, alternative to the photon decay, 
seen as `photoelectric effect' by absorption of the incident photon. 
These microscopic processes will generate `back reactions' leading to 
screening and damping of an incident photon field. 
Notice that these effects are embedded in the equation 
as modification of the external photon momenta $\rp$ and $\rt$ 
through the implicit dependences on the dielectric constant 
(see Eqs.~(\ref{rp_LLL}) and (\ref{rt_LLL})). 
Therefore, what is necessary for examining these effects is, 
in technical terms, to solve Eq.~(\ref{eq:eps_LLL}) self-consistently 
with respect to the dielectric constant appearing on the both sides.



\section{Self-consistent description of dielectric constant 
in the LLL approximation} 
\label{sec:self_consistent}


As summarized in Sec.~\ref{Necessity}, we have observed that 
the naive prescription in obtaining the dielectric constant 
may cause the deviation between the dielectric constant from the 
LLL approximation 
and the preceding result from the numerical integration 
as the photon energy approaches the threshold $\rp=1$ from below, 
and also that there is a contradiction between the limiting value of 
the dielectric constant at the threshold (\ref{eq:eps_limit}) and 
the behavior of the squared photon momentum (\ref{eq:rp_lim}). 
It will turn out, as already suggested in Sec.~\ref{Necessity}, 
that these two problems originate from ``inconsistent" 
treatment of Eq.~(\ref{eq:eps_LLL}) with respect to the dielectric 
constant $\epsilon_\parallel^{\rm LLL}$, and that they are not present 
in fully self-consistent description of Eq.~(\ref{eq:eps_LLL}). 
We will see first in the region below the lowest threshold 
that the self-consistent treatment indeed resolves the devaiation, 
and then will proceed to the region above the threshold 
where the method of numerical integration is not valid. 
In this section, in order to be convinced of the importance of 
fully self-consistent description,
we increase step by step the levels of accuracy in treating 
Eq.~(\ref{eq:eps_LLL}), eventually leading to
the completely self-consistent solution.


\subsection{Three steps towards fully self-consistent description}

\label{subsec:sols}

As discussed in Sec.~\ref{Necessity}, 
Eq.~(\ref{eq:eps_LLL}) contains nontrivial effects 
such as screening and damping of an incident photon field, 
and they are correctly described in a self-consistent description. 
In order to entangle these effects and understand physics 
consequences arising 
from them, 
we solve Eq.~(\ref{eq:eps_LLL}) at three different levels of accuracy 
(I), (II) and (III) in treating the coefficient $\chi_1^{\rm LLL}$. 
We also compare the results of these three methods with the result 
from numerical integration (Num.) which is valid only below the 
threshold.

\begin{enumerate}
\item[]\hspace{-7mm}{\bf (I) ``inconsistent" solution}: 
Instead of solving Eq.~(\ref{eq:eps_LLL}) 
with respect to $\epsilon_\parallel^{\rm LLL}$, we just assume 
$\epsilon_\parallel^{\rm LLL}=1$ in the coefficient function 
$\chi_1^{\rm LLL}$ 
on the right-hand side. This should be a good approximation when the 
deviation of $\epsilon_\parallel^{\rm LLL}$ from unity is small enough. 
The result of this treatment was already shown in Fig.~\ref{fig:eps0}.

\item[]\hspace{-7mm}{\bf (II) partially self-consistent solution}: 
We allow for a real part of 
$\epsilon_\parallel^{\rm LLL}$ on the right-hand side of 
Eq.~(\ref{eq:eps_LLL}). Namely, the equation for the real part 
Re$[\epsilon_\parallel^{\rm LLL}]$ is solved self-consistently, but 
the imaginary part Im$[\epsilon_\parallel^{\rm LLL}]$ is just given 
with the scalar function $\chi_1^{\rm LLL}$ in which 
$\epsilon_\parallel^{\rm LLL}$ is replaced by the solution of 
the real part. 
Thus, this treatment gives a partially self-consistent 
solution and should be a good approximation when the imaginary 
part is small enough. In particular, this treatment is valid 
below the threshold where $\epsilon_\parallel^{\rm LLL}$ does 
not have an imaginary part.

\item[]\hspace{-7mm}{\bf (III) fully self-consistent solution}: 
We treat $\epsilon_\parallel^{\rm LLL}$ in $\chi_1^{\rm LLL}$
as a complex number, 
and solve the real and imaginary parts of Eq.~(\ref{eq:eps_LLL}) fully 
self-consistently to find the dependence of $\epsilon_\parallel^{\rm LLL}$ on
a photon energy. This is the most accurate 
description of the dielectric constant in the LLL approximation. 

\item[]\hspace{-7mm}{\bf (Num.) self-consistent numerical solution 
below the threshold}: As already mentioned in the previous sections, one can 
directly evaluate the scalar functions $\chi_i$ ($i=0,1,2$) by numerical 
integration as long as 
the photon momentum is below the threshold $\rp<1$. 
One can also solve Eqs.~(\ref{epsilon_perp}) and (\ref{epsilon_parallel}) 
self-consistently. However, since this is possible only below the threshold,
the dielectric constants are real. This treatment should be compared with 
(II) and (III) below the threshold. 

\end{enumerate}

\subsection{Threshold condition in self-consistent description}
\label{subsec:threshold}

Before we present results for the dielectric constant in the 
self-consistent description, let us revisit the threshold condition $\rp=1$.  
We discussed that the threshold condition $\rp = 1$ gives the threshold 
photon energy $\ome^2_{\rm th} 
= 1/ (1-\cos^2\theta)$ (see Fig.~\ref{fig:eps0}), 
if $\epara=1$ is assumed in $\rp$. In this case, 
the region below the threshold is simply represented by 
$\ome^2 < \ome^2_{\rm th} $ 
as shown in Fig.~\ref{fig:eps0} (where $\ome^2_{\rm th}=2$ 
for $\theta=\pi/4$). 
However, it should be noticed first that, 
if we keep the real part of $\epara$ in $\rp$, the threshold photon energy 
is given in an $\epara$-dependent form as, 
\begin{eqnarray}
\ome^2_{\rm th} = \frac{1}{ 1 - \epara \cos^2 \theta  }
\label{eq:ome_thr}
\ \ ,
\end{eqnarray}
and this condition is inversely solved as, 
\begin{eqnarray}
\epsilon_\parallel = \frac{ 1 }{ \cos^2 \theta } 
\left(1 - \frac{1}{\ome^2_{\rm th}} \right)
\label{eq:boundary}
\ \ .
\end{eqnarray}
The above relation (\ref{eq:boundary}) separates the kinematical regions 
below and above the threshold of photon decay. 
Therefore, in the $\epsilon_\parallel$-$\ome^2$ plane, 
the threshold line is not just a vertical line 
(as shown in Fig.~\ref{fig:eps0}), but is a curve given 
by Eq.~(\ref{eq:boundary}). 
Inserting the limiting value $\epslim$ (\ref{eq:eps_limit}) 
into Eq.~(\ref{eq:ome_thr}), we find that the photon energy 
approaches infinity, $\ome \rightarrow \infty$, 
as the dielectric constant approaches the limiting value. 
Equivalently, Eq.~(\ref{eq:boundary}) shows that 
the dielectric constant does not reach the limiting value 
$\epslim$ at $\rp(\epara)=1$, 
as long as the photon energy $\ome$ is finite. 
This observation implies that the dielectric constant could stay 
in a real value when $\ome$ increases from zero to infinity, 
and thus that the energy dependence of the dielectric constant would be 
significantly modified from the one in Fig.~\ref{fig:eps0} 
obtained without any self-consistent treatment. 


\begin{figure}[t]
 \begin{minipage}[t]{0.66\hsize}
  \begin{center}
   \includegraphics[width=\hsize]{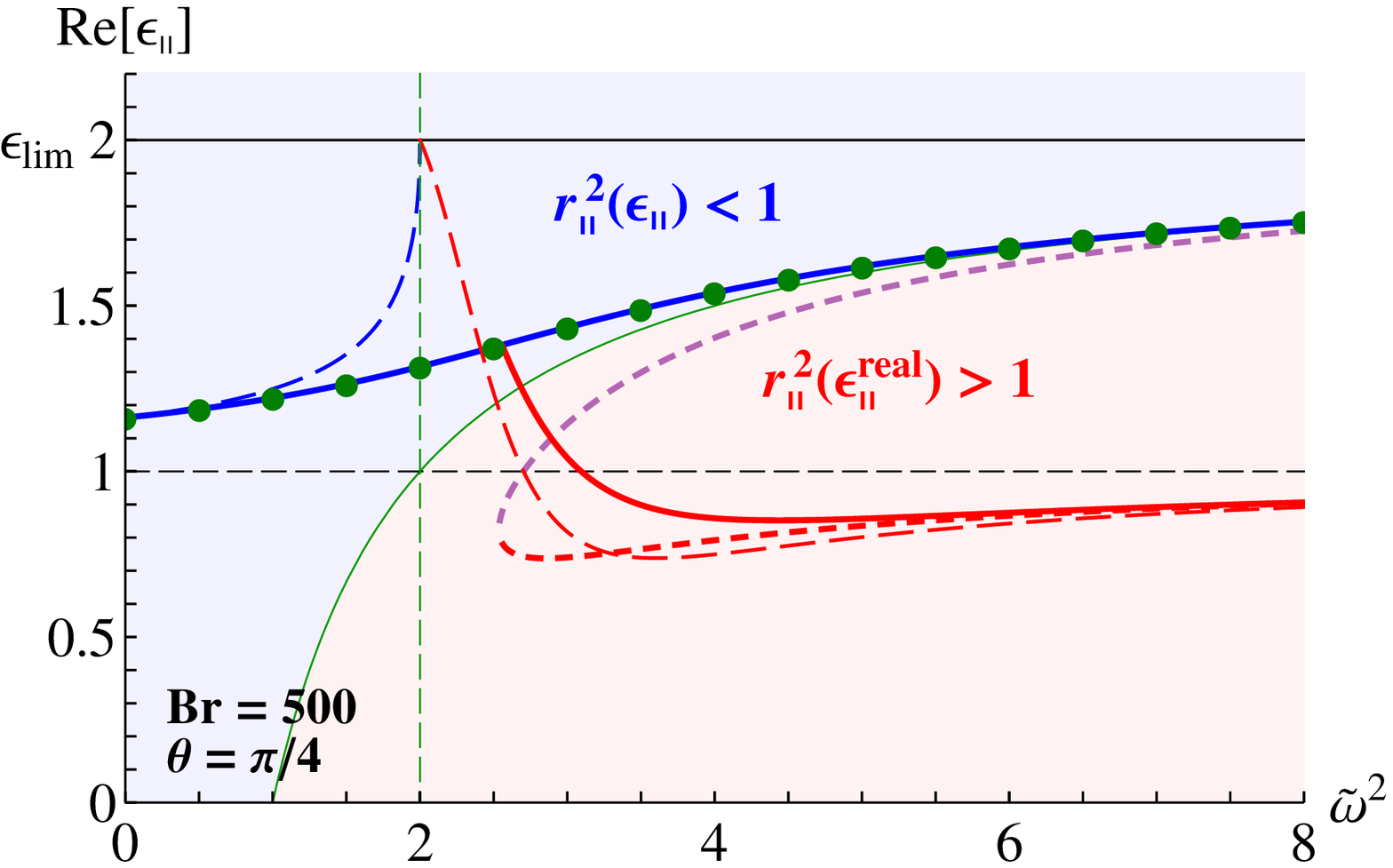}
  \end{center}
 \end{minipage}
 \begin{minipage}[t]{0.65\hsize}
  \begin{center}\hspace*{2.mm}
   \includegraphics[width=\hsize]{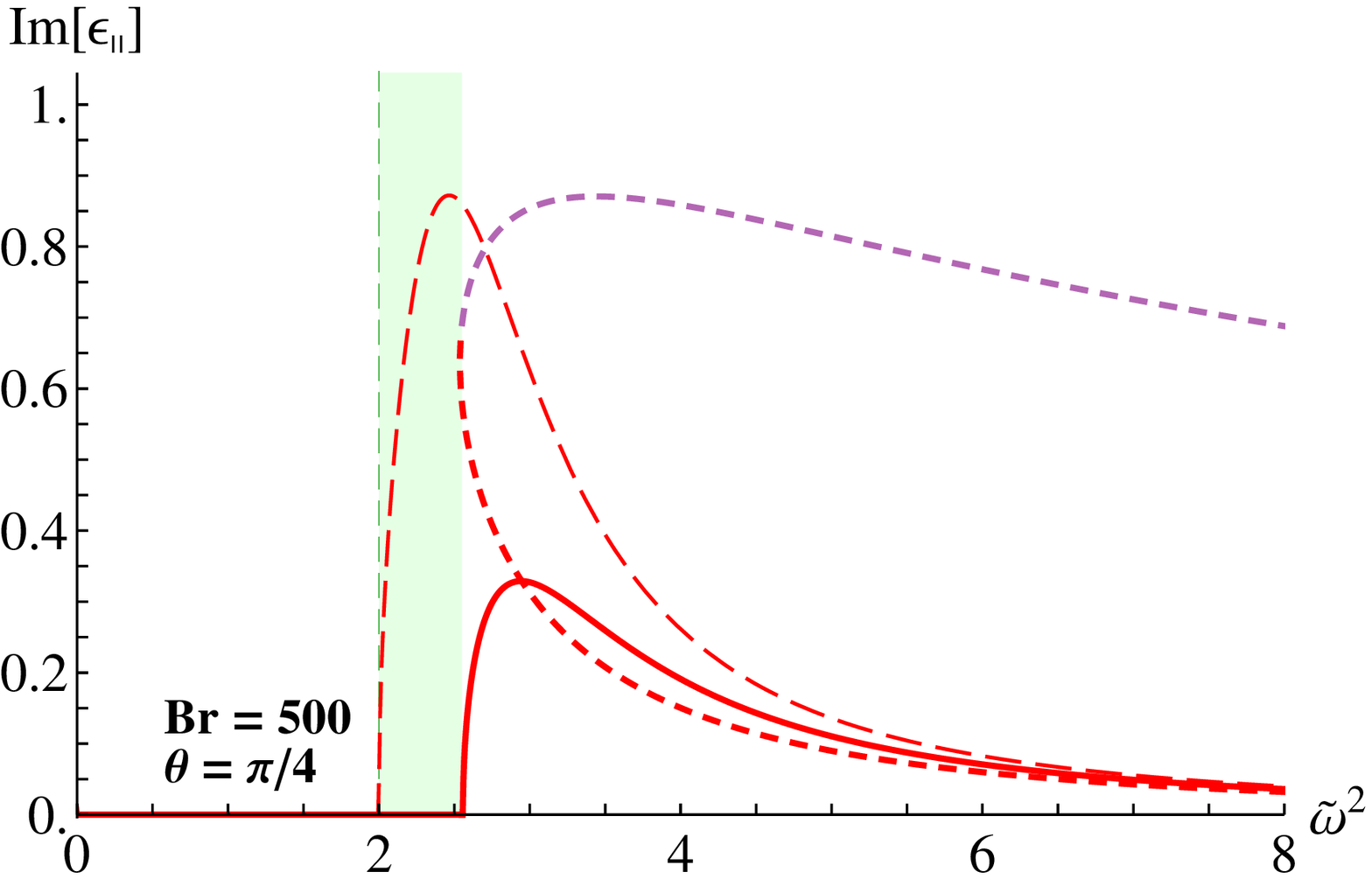}
  \end{center}
 \end{minipage}
\begin{center}
   \includegraphics[width=\hsize]{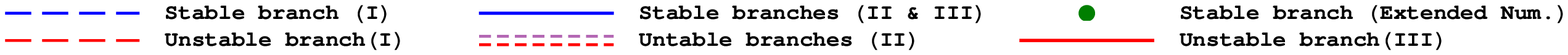}
\end{center}\vspace*{-3mm}
\caption{
Dielectric constant for $\theta=\pi/4$ and $\Br = 500$: 
a shift of the threshold in solution (III) 
from that in solution (I) is indicated with a green stripe in the lower panel. 
}
  \label{fig:full500}
\end{figure}


\subsection{Back reactions in complex dielectric constant}

\label{subsec:energy_dep}

Now let us finally show results from the self-consistent description. 
Figure~\ref{fig:full500} is a compilation of all the results of the real 
and imaginary parts of the dielectric constant obtained with four 
different treatments from (I) to (III) and (Num.). Upper and lower panels 
show the real and imaginary parts, respectively. We have taken the same values 
for $\Br$ and $\theta$ as in Fig.~\ref{fig:eps0}, namely  $\Br = 500$ 
and $\theta=\pi/4$. The threshold boundaries specified by 
Eq.~(\ref{eq:boundary}) and $\rp (\epara=1) = 1$ are indicated 
with green solid and dashed lines, left (right)
of which is filled with light blue (pale red) and corresponds to the 
region below (above) the threshold. 
Long dashed lines (blue and red) are the ``inconsistent" 
solutions from (I), 
short dashed lines (purple and red) are the partially self-consistent
solutions from (II), and finally 
thick bold lines (blue and red) are the fully self-consistent solutions 
from (III). These are compared with green dots which are the results of 
numerical integration (Num.). Notice that this is the same result as shown 
in Fig.~\ref{fig:eps0}, but is now extended to larger $\ome^2$ because the 
threshold line is modified from a vertical line to a curve and the region 
below the threshold (filled with light blue color) extends to 
infinite $\ome^2$.

\subsubsection{Screening of incident light}

\label{subsubsec:screen}

Let us first consider the change from the ``inconsistent" solution (I)
shown as blue and red long-dashed lines, 
to the partially self-consistent solution (II) shown as 
a blue solid line and purple and red short-dashed lines. 
Look at the blue solid line in the upper panel, which is a common 
solution in treatments (II) and (III). Also, it coincides with the 
green dots which are from the numerical integration (Num.). 
Thus, as expected, the treatment (II) already gives a final result 
in this region below the threshold. 
Since this line is not accompanied by an imaginary part at any 
photon energy, we call this {\it stable branch}. 
In particular, we now understand that the origin of the 
deviation found  in Fig.~\ref{fig:eps0} 
between the results from the LLL approximation with the 
naive prescription (blue long-dashed line) and the numerical 
integration (green dots) is the incomplete treatment of 
the equation (\ref{eq:eps_LLL}), 
and that the value of the dielectric constant in the naive 
prescription is reduced (i.e., screened) due to back reactions. 
In addition to the stable branch, 
we found other 
two solutions (purple and red short-dashed lines) in the 
region, $\rp(\epara) > 1$. 
Namely, in this partially self-consistent treatment (II), 
we obtained the dielectric constant as a three-valued function of 
the photon energy $\ome$. 
The latter two branches in the region, $\rp(\epara) > 1$, 
smoothly connects to each other. 
In contrast to the solution in the region, $\rp(\epara) < 1$, 
these branches are accompanied by imaginary parts 
as shown in the lower panel, 
and thus we call them {\it unstable branches}.

As described above, the green solid line divides the kinematical regions 
$\rp(\epara) < 1$ and $\rp(\epara) > 1$, 
while the green dashed line $\rp(\epara=1) < 1$ and $\rp(\epara=1) > 1$. 
Noting a correspondence between these lines, we find that the partially 
self-consistent solution (II) still maintains 
a topological structure of (I) as correspondences 
between stable branches and unstable branches, respectively: 
the blue long-dashed (red long-dashed) line is modified to 
the blue solid (purple and red short-dashed) line in the solution (II). 
Location of the connecting point between stable and unstable branches, 
where the dielectric constant takes the limiting value $\epslim$, 
has moved away from $\ome^2 = 1/(1-\cos^2\theta)$ in (I) 
to $\ome^2 = \infty$ in (II). 

The lower panel in Fig.~\ref{fig:full500} shows an imaginary part of the dielectric constant 
associated with the unstable branches shown in the upper panel. 
Since the partially self-consistent solution (II) gives the double-valued unstable branch, 
an associated imaginary part is also a double-valued function of the photon energy. 
As shown with a green stripe in the background, 
the threshold of the photon decay shifts upward because of the 
modification of the real part incorporated in $\rp( \epara )$.


\subsubsection{Damping of incident light}

\label{subsubsec:damp}

Now let us see the change of solutions from (II) to (III). In partially 
self-consistent treatment (II), we treated the dielectric constant in 
$\chi_1^{\rm LLL}$ as a real value. However, since the unstable branch 
has a large imaginary part (see purple short-dashed line in 
Fig.~\ref{fig:full500}), we should not assume that the dielectric 
constants in $\rp$ and $\rt$ are real values. 
Physically, this indicates that we have to self-consistently take 
into account the damping of an incident photon due to decay into a 
fermion-antifermion pair in 
the external magnetic field when the decay rate is sizable.

We thus solve simultaneously the equations for real and imaginary parts 
of Eq.~(\ref{eq:eps_LLL}) without any assumption. This is the result of 
treatment (III). Maintaining both the real and imaginary parts of the 
dielectric constant in the photon momenta $\rp$ and $\rt$, 
we find that the structure of the branch in the kinematical region above the 
threshold is significantly modified. 
The most remarkable modification of the structure is that 
an unstable branch in the real part, indicated with a red solid line, 
directly connects to the stable branch below the threshold. 
According to this, the dielectric constant reduces 
from the three-valued function to a double-valued function of the photon energy. 
This edge of the unstable branch close to the connecting point is still 
accompanied by an imaginary part as shown in the lower panel, and thus 
is a state unstable to the decay. 
We find that a magnitude of the imaginary part is diminished in 
(III) compared to those of the imaginary parts obtained in the 
other two cases, (I) and (II). This result can be intuitively understood 
as follows: the magnitude of the imaginary part should be suppressed 
in the self-consistent treatment, 
since the amplitude of the incident photon damps due to the decay. 
Indeed, the amplitude of the photon propagation is given by the complex 
refractive index related to the dielectric constant as shown in 
Eqs.~(\ref{eq:photon_field}). 
In Sec.~\ref{sec:refractive_index}, we will also show 
a complex refractive index obtained from 
the dielectric constant shown in this section.

\begin{figure}[t]
 \begin{minipage}[t]{0.655\hsize}
  \begin{center}
   \includegraphics[width=\hsize]{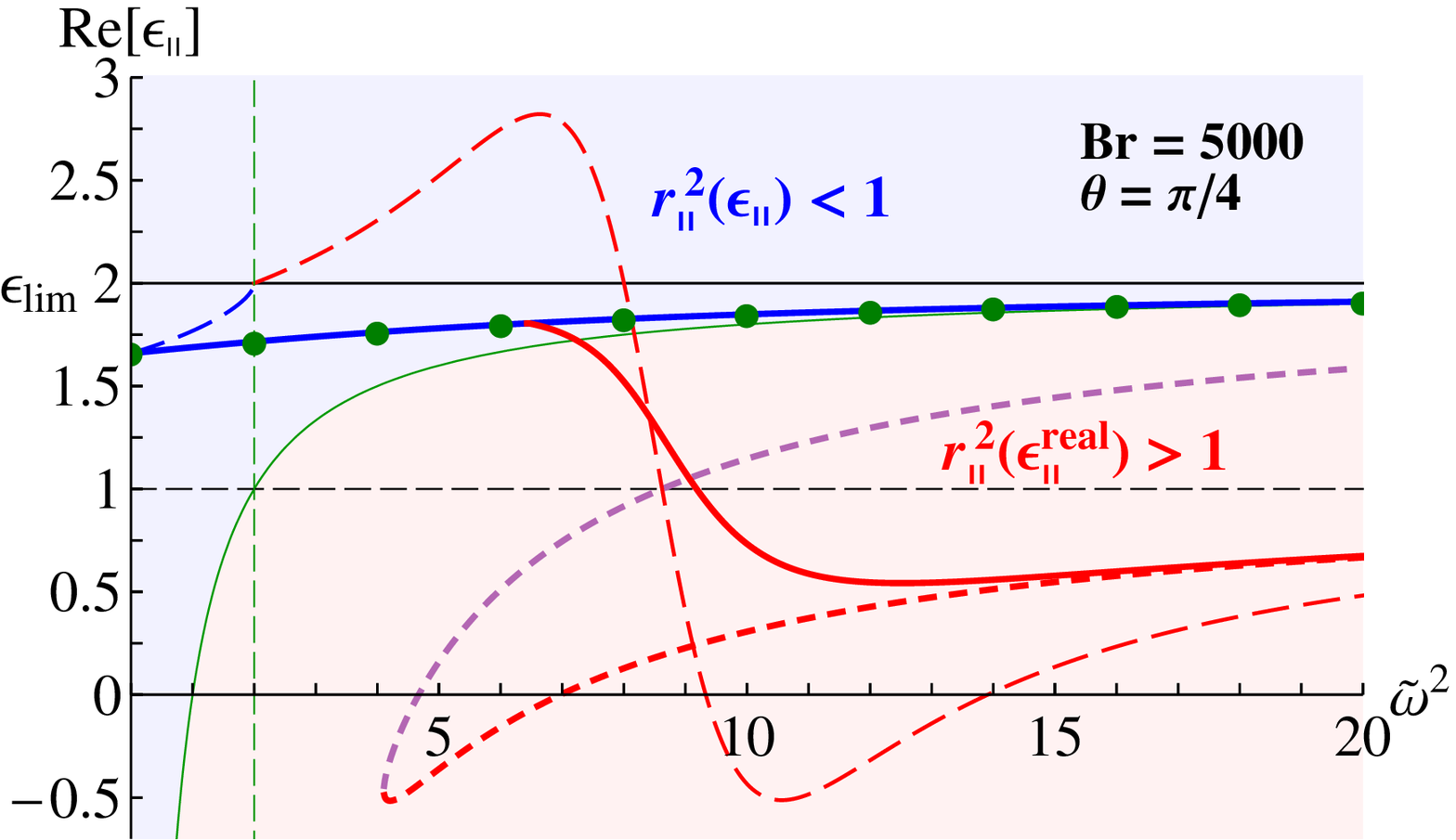}
  \end{center}
 \end{minipage}
 \begin{minipage}[t]{0.645\hsize}
  \begin{center}\hspace*{2.mm}
   \includegraphics[width=\hsize]{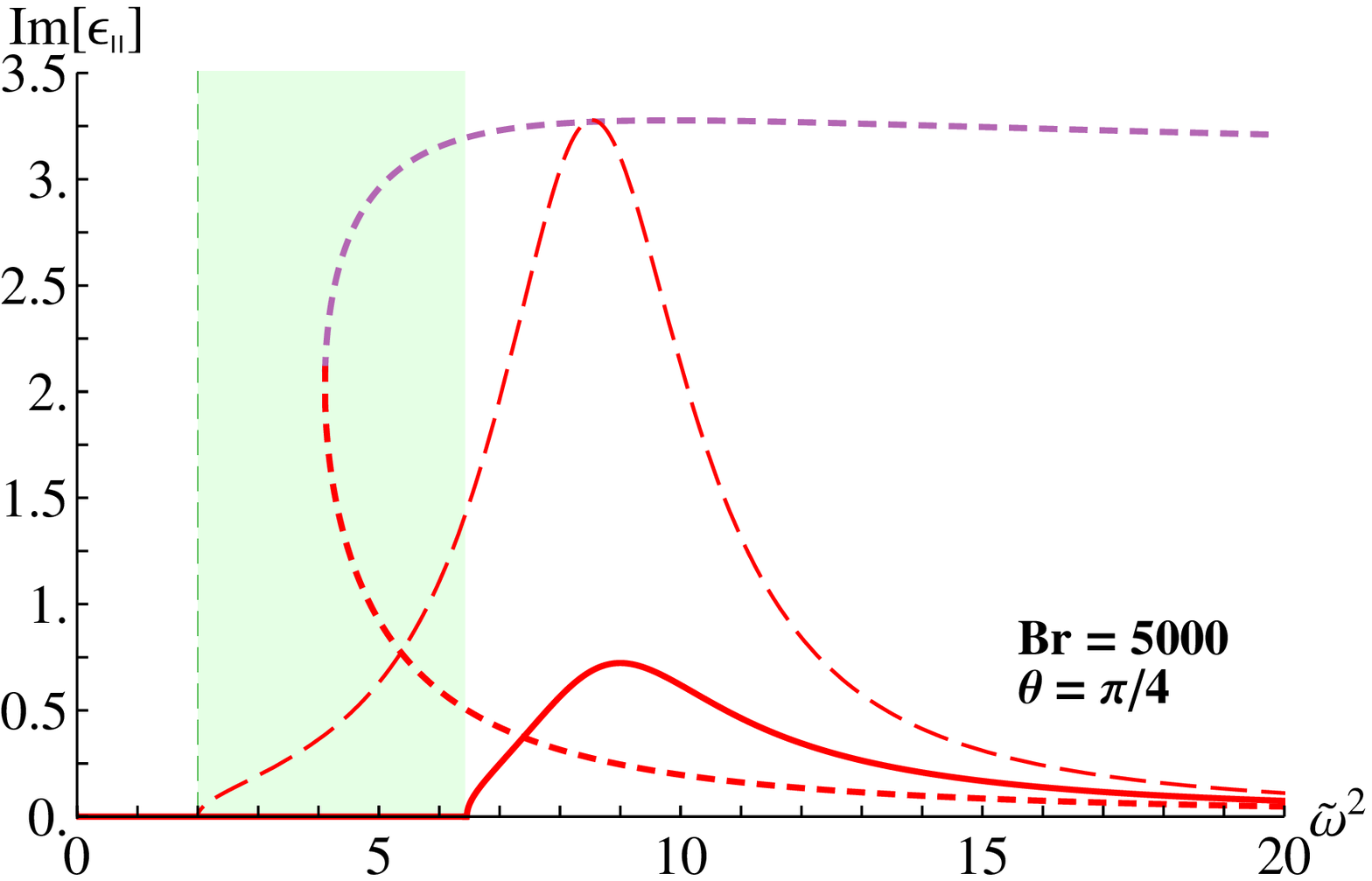}
  \end{center}
 \end{minipage}
\caption{
Dielectric constant for $\theta=\pi/4$ and very strong magnetic field 
$\Br = 5000$: each line shows the corresponding quantities in the 
legend of Fig.~\ref{fig:full500}. 
}
  \label{fig:full5000}
\end{figure}

\subsection{Results in stronger magnetic field}

We shall see a dielectric constant in a stronger magnetic field. 
Figure~\ref{fig:full5000} shows the dielectric constant at much larger 
value of $\Br = 5000$. 
Each line shows the corresponding quantities indicated in the 
legend of Fig.~\ref{fig:full500}. Compared to the results for $\Br=500$ 
in Fig.~\ref{fig:full500}, one finds that the modification of the real 
part in (I) is larger, in particular in $\ome^2>2$, with a nontrivial wavy 
behavior. However, as shown in red solid line of treatment (III), 
it is suppressed if we self-consistently
incorporate a screening effect on the incident photon field 
caused by the induced vacuum polarization. 
Consequently, the real part eventually behaves similarly as
in the previous case with $\Br = 500$, but its magnitude shows 
moderate enhancement. 
A significant effect of the self-consistent treatment is also 
found in the imaginary part. 
Whereas magnitudes of the imaginary parts in (I) and (II) are much 
larger than those in Fig.~\ref{fig:full500}, 
they are suppressed in (III) to give moderate enhancement 
of the imaginary part at $\Br=5000$, as in the real part. 
These results indicate that 
stronger distortion of the Dirac sea induces 
stronger effects of the competing back reactions. 
As for the energy dependence, the stable branch extends up to 
infinite photon energy as in Fig.~\ref{fig:full500}. 
The connecting point between the stable and unstable branches 
moves to a larger photon energy compared to the case in 
Fig.~\ref{fig:full500}, 
indicating a larger upward shift of the threshold. 
The imaginary part has a broader profile in Fig.~\ref{fig:full5000}. 
We will investigate $\Br$-dependence more systematically for 
the refractive index in Sec.~\ref{subsec:Br} 
and the dielectric constant in Appendix~\ref{sec:eps_Br}.


\section{Refractive index in the LLL approximation} 
\label{sec:refractive_index}


Thus far, we have considered only the dielectric constants $\epsilon$.
However, in some cases, it is more convenient to treat the refractive 
indices $n$, which can be immediately obtained from the dielectric constants 
through a simple relation $n^2 = \epsilon$. 
In this section, we will present results for the refractive index in the LLL 
approximation, and in particular we will focus on its dependences on 
the angle and magnetic field strength. 

When the dielectric constants have imaginary parts, 
$\epsilon = \epsilon_{\rm real} + i \, \epsilon_{\rm imag}$, 
we can similarly define real and imaginary parts of the refractive 
indices as $n = n_{\rm real} + i \, n_{\rm imag},$ 
which are explicitly given with respect to the refractive index as, 
\begin{eqnarray}
\label{eq:ref_real}
&& n_{\rm real} \ = 
\ \frac{1}{\sqrt{2}} \sqrt{  | \epsilon | + \epsilon_{\rm real}  }\, ,  
\\
&& n_{\rm imag} \ = 
\ \frac{1}{\sqrt{2}} \sqrt{  | \epsilon | - \epsilon_{\rm real}  }  
\, ,
\label{eq:ref_imag}
\end{eqnarray}
with the magnitude $| \epsilon |$ being defined by 
$| \epsilon |=\sqrt{ \epsilon_{\rm real} ^2 + \epsilon_{\rm imag} ^2 }$.


\begin{figure}[t]
 \begin{minipage}[t]{0.68\hsize}
  \begin{center}\hspace{-6mm}
   \includegraphics[width=\hsize]{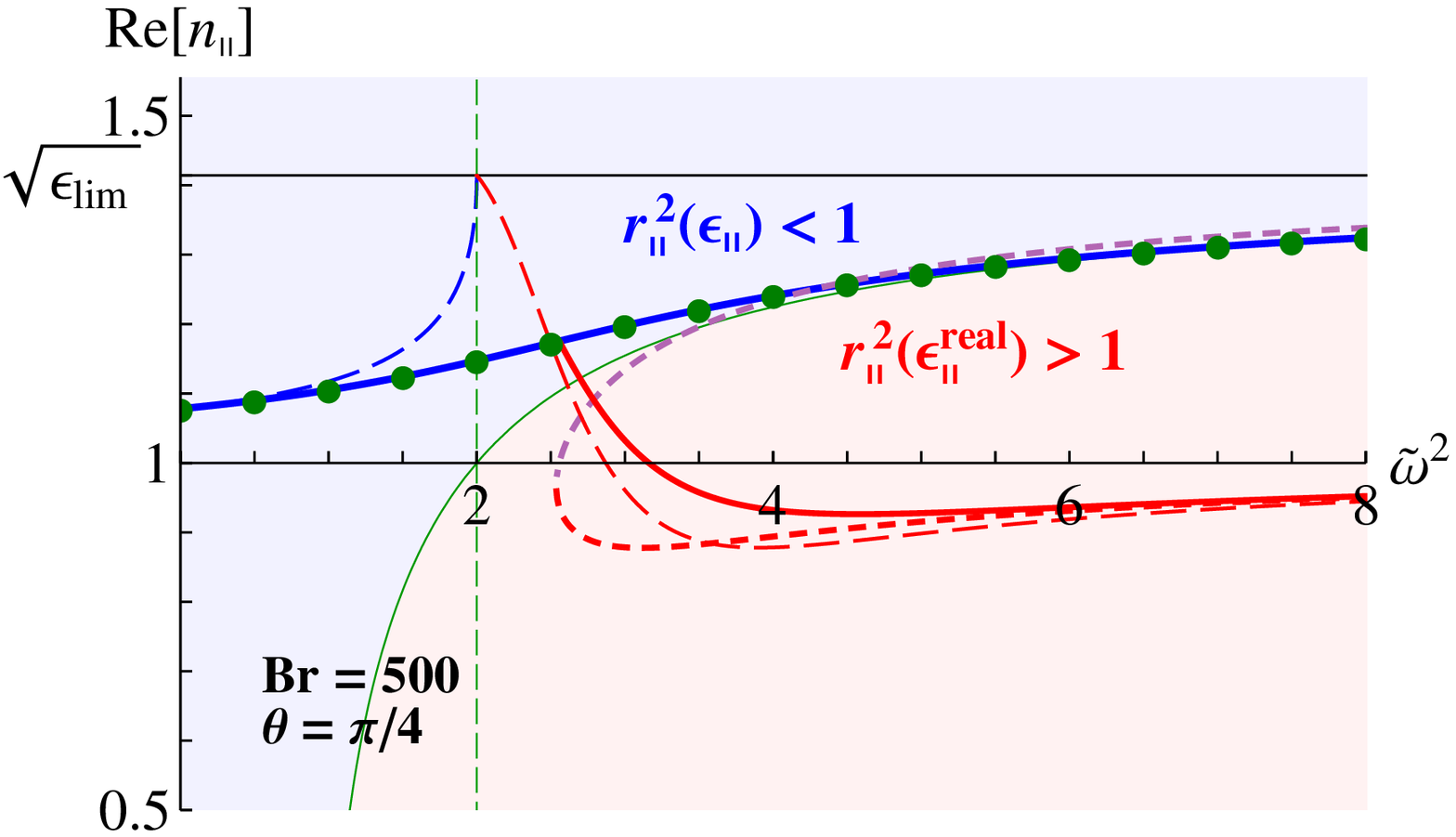}
  \end{center}
 \end{minipage}
 \begin{minipage}[t]{0.63\hsize}
  \begin{center}\hspace*{1.5mm}
   \includegraphics[width=\hsize]{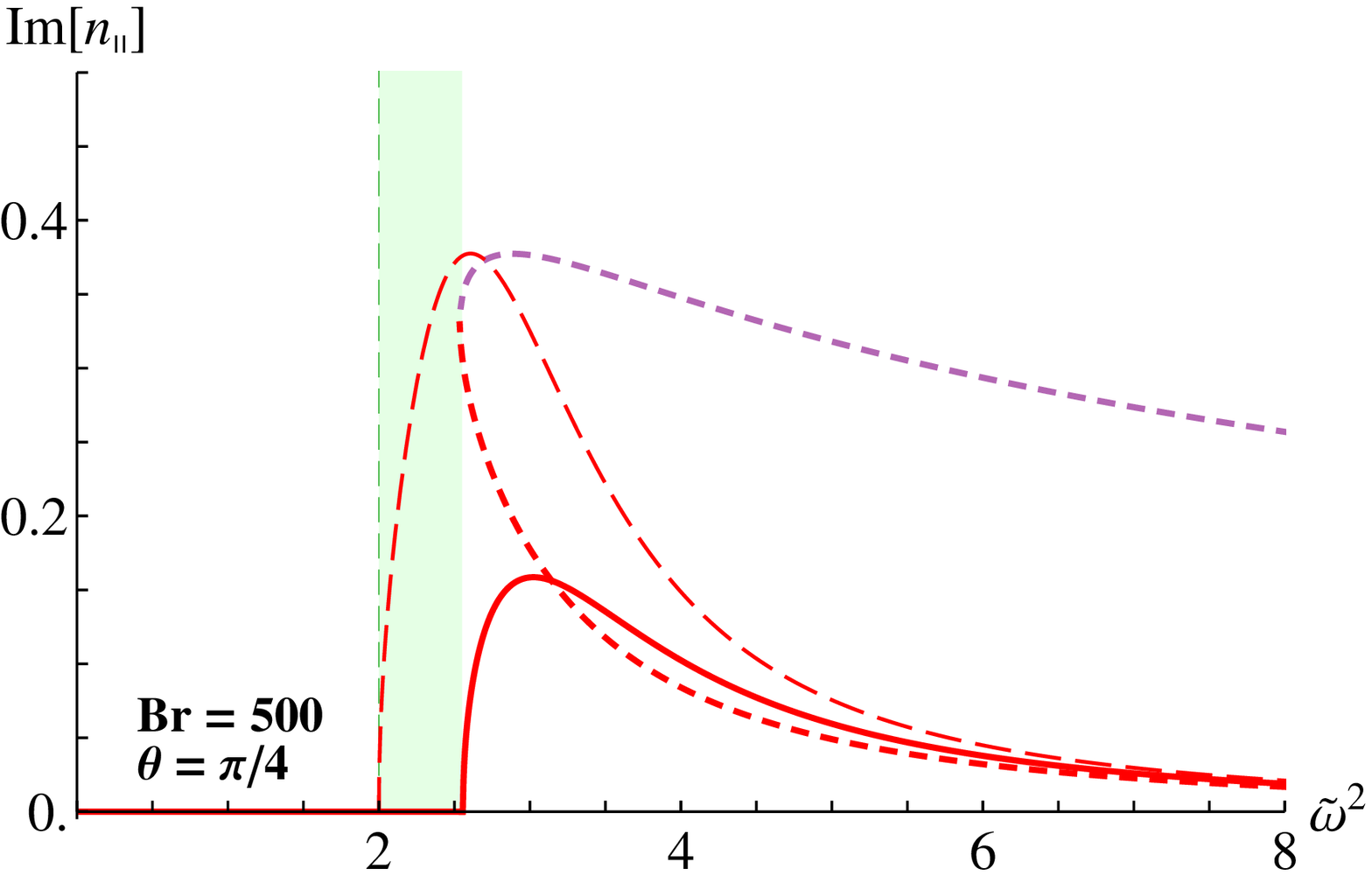}
  \end{center}
 \end{minipage}
\vspace{-0.05cm}
\caption{
Refractive index $n_\parallel$ for $\theta=\pi/4$ and $\Br = 500$: 
each line shows the corresponding quantities explained in the
 legend in Fig.~\ref{fig:full500}. 
}
  \label{fig:ref500}
\end{figure}


As we saw in the previous sections,  only the parallel mode 
of the dielectric constant $\epara$ remains nontrivial in the LLL 
approximation, and the same is true for the refraction index: 
we have only $n_\parallel$ as a nontrivial mode. 
Figure~\ref{fig:ref500} shows the refractive index $n_\parallel$ at $\Br =500$ 
as a function of (scaled) photon energy squared $\ome^2$ 
that corresponds to the dielectric constant in Fig.~\ref{fig:full500}. 
We find that the refractive index has a similar structure in 
photon energy dependence, and that it becomes complex when the 
dielectric constant takes the unstable branch. 
We do not repeat the explanation for the transition of the solutions 
from the naive prescription (I) to the fully self-consistent 
treatment (III). All the 
explanations for the dielectric constant $\epsilon_\parallel$
equally apply to the refractive index $n_\parallel$. In this section,
we rather discuss dependences of the self-consistent 
solutions on the propagation angle $\theta$ and the magnetic 
field strength $\Br$. In fact, the same analyses were in advance 
performed for the dielectric constant $\epara$ as summarized in 
Appendix~\ref{sec:eps_Br}, and we have used those results 
to compute the refractive index through the relations 
(\ref{eq:ref_real}) and (\ref{eq:ref_imag}).


\subsection{Dependence on propagation angle}
\label{subsec:angle}

\begin{figure}[t]
  \begin{center}
		\includegraphics[width=0.64\hsize]{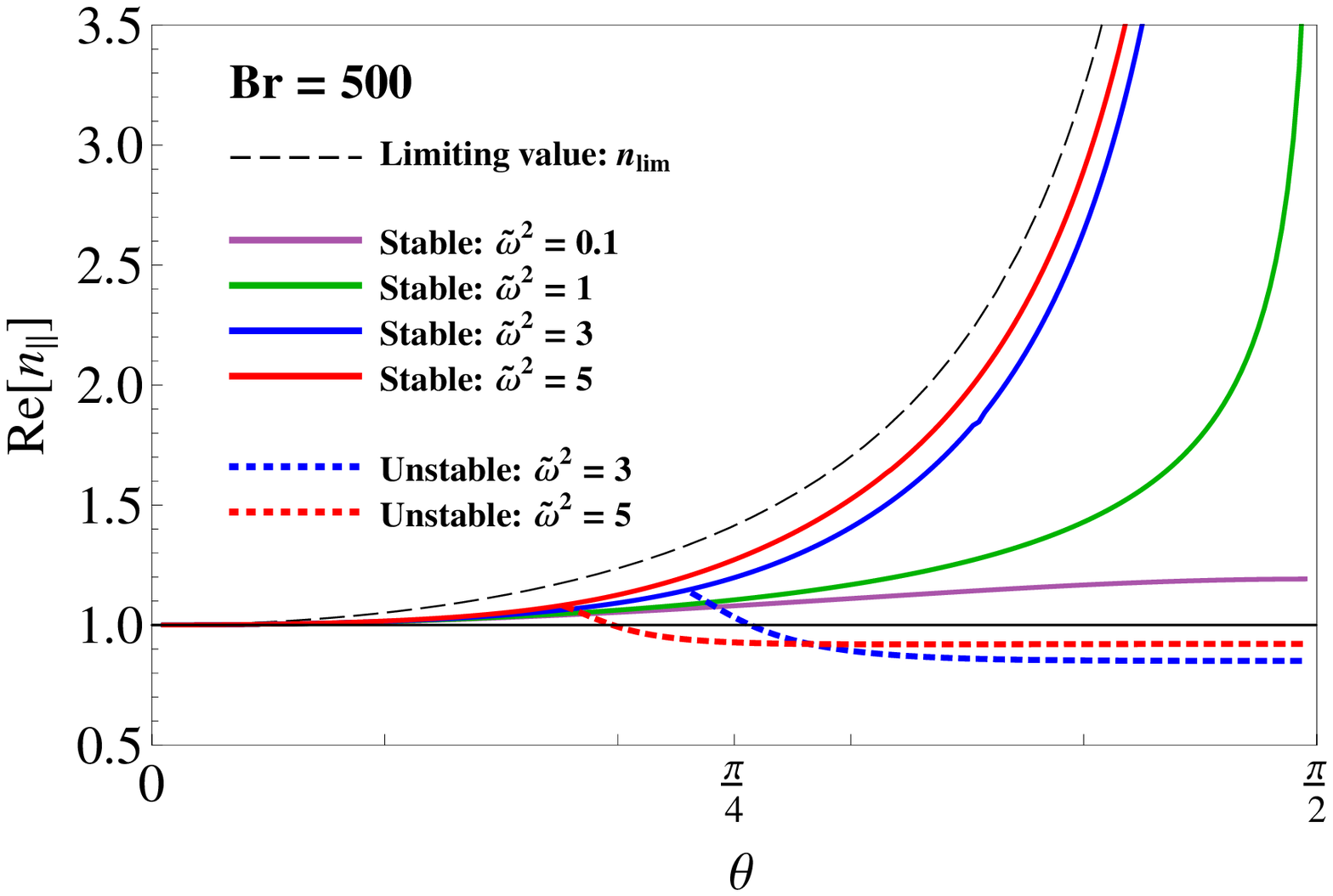}
\hspace*{-0.7cm}
		\includegraphics[width=0.67\hsize]{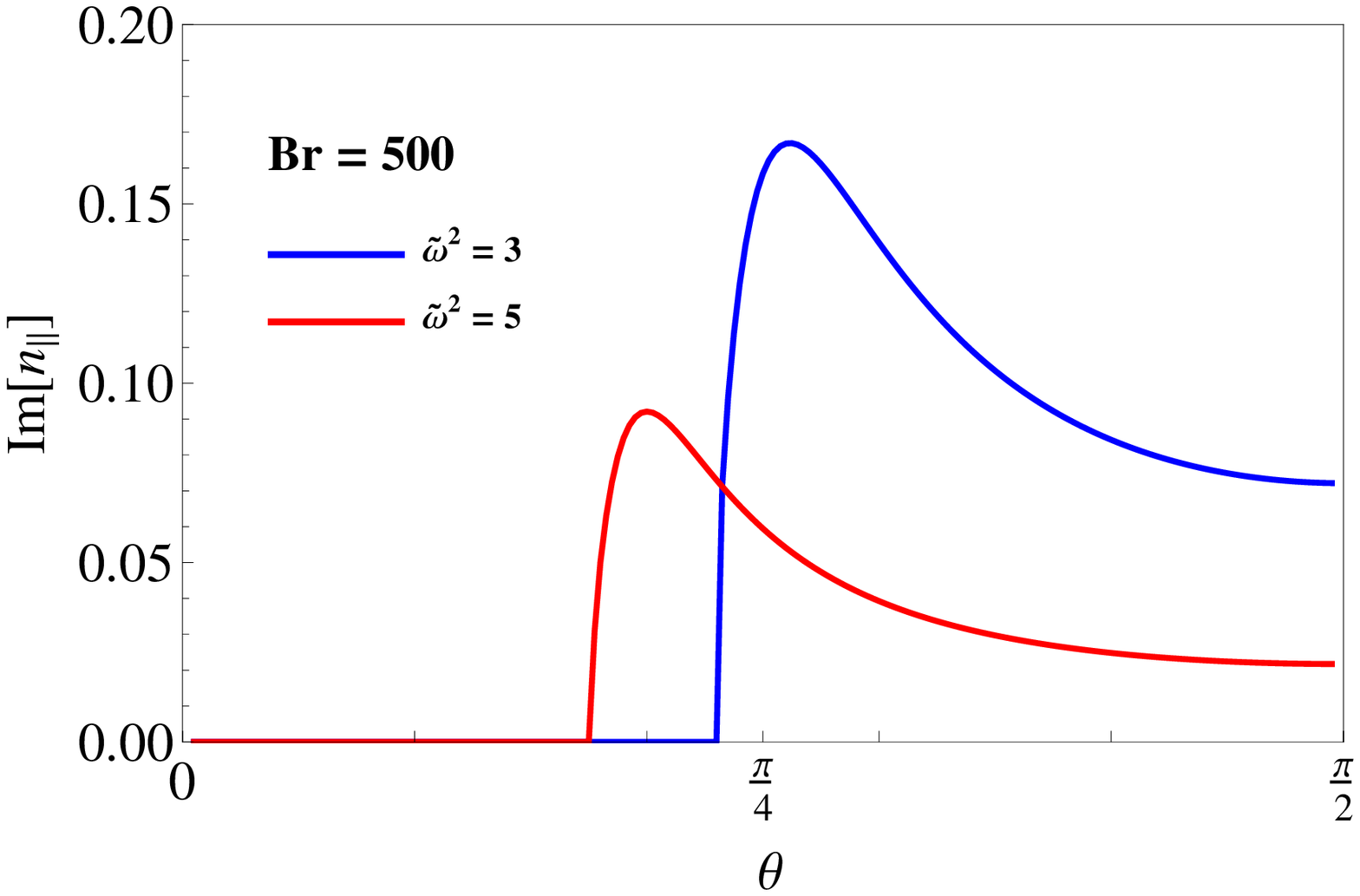}
	\end{center}
	\vspace{-0.5cm}
\caption{
Refractive index with varying propagation angle $\theta$: 
The refractive index takes the same value, 
when photon propagation is oriented in the directions, $\pm\theta, \ \pi \pm \theta$. 
}
\label{fig:angle}
\end{figure}

In the previous section and in Fig.~\ref{fig:ref500}, the results of $\epsilon_\parallel$ and $n_\parallel$ were shown  
only for the angle $\theta=\pi/4$. However,
there is a significant angle-dependence in the results, which we are going to 
present in this subsection. 
Upper and lower panels in Fig.~\ref{fig:angle} 
represent real and imaginary parts of the complex refractive index 
when the propagation angle varies from $\theta = 0$ to $\pi/2$. 
We have shown the results only in this region of $\theta$ 
because the refractive index has the following symmetry: 
$\epara(\theta)=\epara(-\theta)=\epara(\pi \pm \theta)$. 
In the upper panel, solid lines are stable branches with 
different photon energies. A long-dashed line indicates 
a limiting value $n_{\rm lim}$ of the refractive index on the stable 
branch. It is given by 
$n_{\rm lim}=\sqrt{\epsilon_{\rm lim}} = \vert \cos\theta \vert^{-1}$, 
namely a square root of the limiting value of the dielectric constant 
(\ref{eq:eps_limit}). 
As seen in Fig.~\ref{fig:ref500} for $\theta = \pi/4$, 
the refractive index approaches the limiting value $n_{\rm lim}$
as the photon energy increases. At a certain photon energy above around 
$\tilde \omega^2 \sim 1$, 
the refractive index on the stable branch strongly depends on 
the angle. The real parts grow with increasing $\theta$, 
indicating that, when a photon propagates almost perpendicularly 
to the magnetic field, the effects of vacuum birefringence appear 
more strongly and photon's phase velocity becomes significantly small. 

While we obtained the simple expression of the limiting value $n_{\rm lim}$ 
within 1-loop accuracy of the vacuum polarization tensor, 
it is still an open question how higher-order diagrams contribute 
to the refractive index. 
Indeed, we notice that the limiting value $n_{\rm lim}$ diverges in particular two cases 
when the photon propagates in the perpendicular directions $\theta = \pm \pi/2$ 
with energy $\tilde \omega^2 = 1$. 
These singular behaviors would suggest that further careful investigation is required 
when the modification of the refractive index is very large, 
e.g., the vacuum polarization would be suppressed by mutual Coulomb interaction 
between a fermion and antifermion pair appearing in the loop part 
which corresponds to 
photon exchanges in the multi-loop diagram.

There are also unstable branches (short-dashed lines) when the angle is large. 
This is clearly seen in the lower panel where the imaginary parts are shown. 
The stable (unstable) branches appear in small (large) $\theta$ region 
because the threshold condition $\rp = 1$ depends on the angle. 
Also, the angle where unstable branch starts to appear depends on the photon 
energy, and decreases with increasing energies. Therefore, if the photon
energy is large enough, it can decay even at small angles. 
Note, however, that the decay never occurs at $\theta = 0$ within 
constant magnetic field case, where the refractive index persists to 
be unity as in the ordinary vacuum.


\begin{figure}[t]
  \begin{center}
   \includegraphics[width=0.75\hsize]{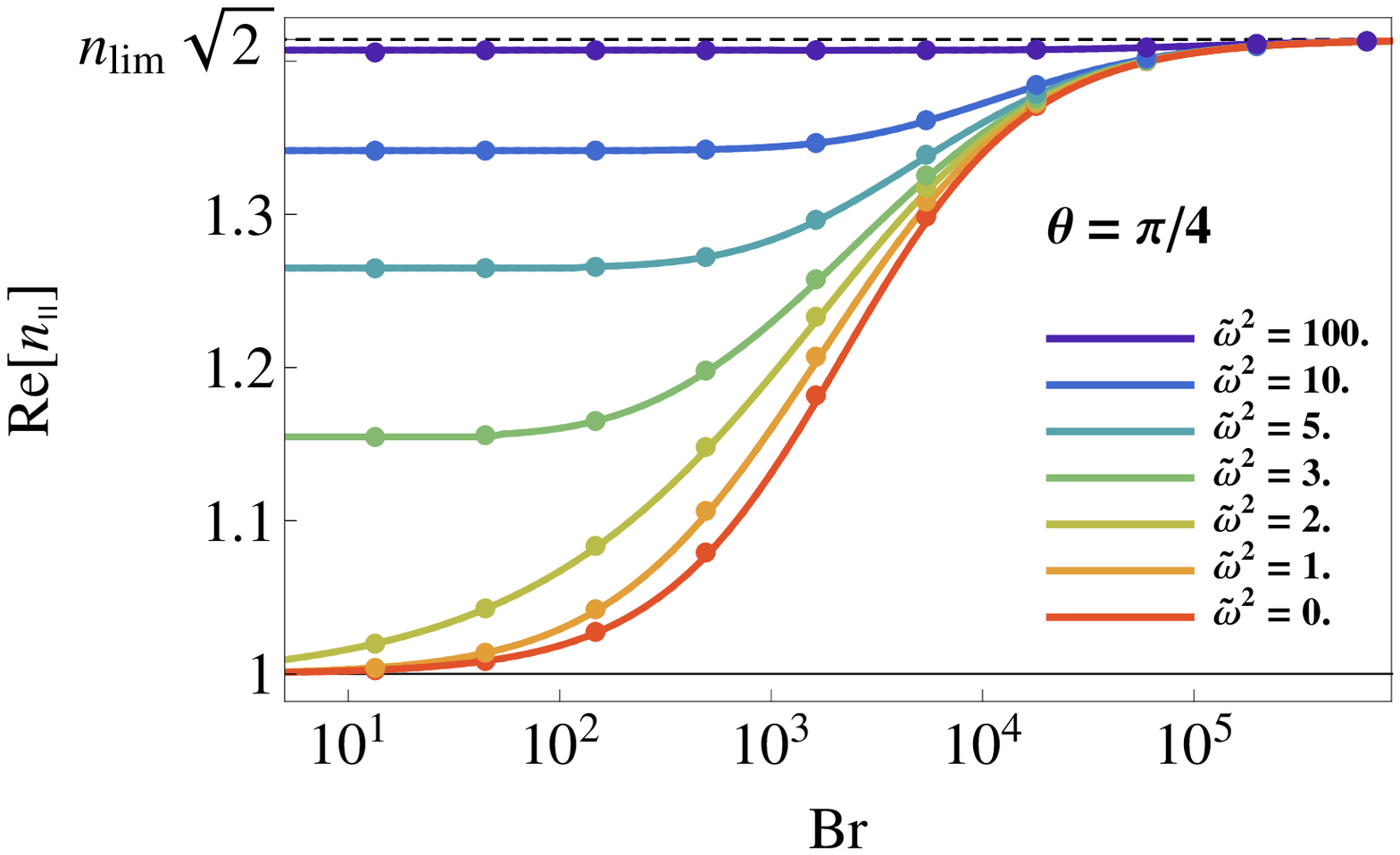}
   \includegraphics[width=0.76\hsize]{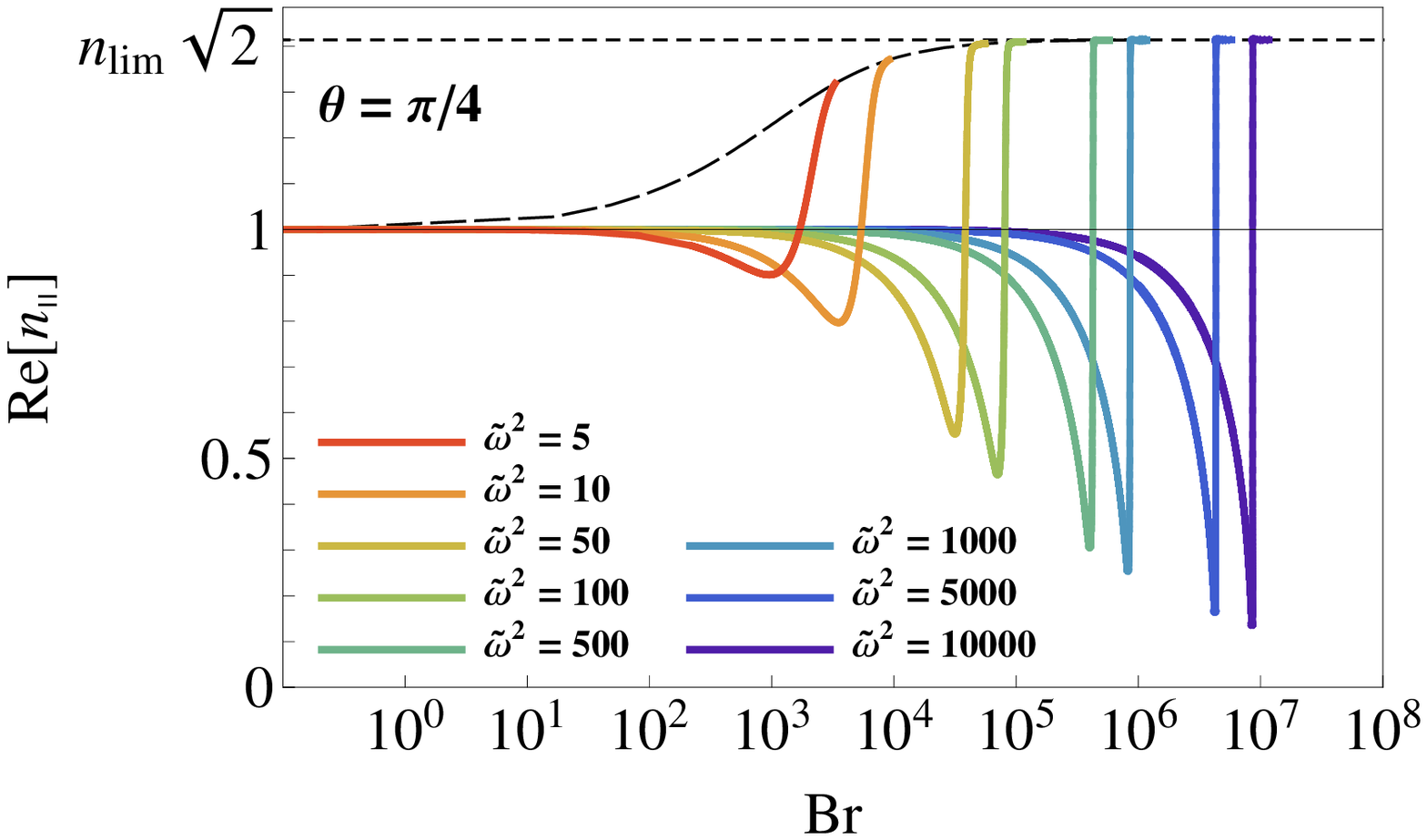}	
  \end{center}
\caption{
$\Br$ dependence of the real part of $n_\parallel$ 
in the LLL approximation. Lines show the refractive index 
at $\theta = \pi/4$ and various photon 
energies. Upper and lower panels show the stable and unstable branches, 
respectively. In the upper panel, results from the LLL approximation 
agree well with those from the numerical integration which are indicated 
with dots. 
}
\label{fig:ref_Br}
\end{figure}


\subsection{Dependence on the magnetic-field strength}

\label{subsec:Br}

In this subsection, we present the dependence of the refractive index on 
the magnitude of the external magnetic field. 
The real part of the refractive index on stable and unstable branches, 
and the imaginary part on the unstable branch are respectively shown below.

\subsubsection{Real part}

Upper panel of Fig.~\ref{fig:ref_Br} shows a $\Br$-dependence of 
the 
refractive index $n_\parallel$ on the stable branch. 
Since the stable branch is in the region below the threshold, the 
refractive index has only a real part. 
Curves and dots indicate the results from the LLL approximation and 
the numerical integration, respectively, 
which agree well with each other in the strong-field limit $\Br \agt 10 $. 
Thus, we again confirm the validity of the LLL approximation
when the magnetic field is strong enough $\Br \agt 10 $. 
We find that, at any photon energy, magnitude of the 
refractive index increases with increasing $\Br$, and saturates 
in the limit of strong magnetic field. The limiting value is 
given by $\nlim = \sqrt{ \epslim }  $ as mentioned in the last subsection. 
While we discussed this limiting value $\nlim $ as reached 
in the limit of a large photon energy (see Eq.~(\ref{eq:boundary})), 
the plot shows the same limit is reached, at any photon energy, 
in the strong-field limit. 
Remind that the limiting value of the dielectric constant $\epslim $ 
is obtained in the limit in which the scalar function $\chi_1$ becomes divergently large $\chi_1 \gg 1$ 
as the photon momentum approaches the threshold $\rp(\epara) \rightarrow 1$. 
Note also that the scalar function (\ref{eq:chi1_LLL}) is proportional to 
the field strength, $\chi_1 \propto \Br$, 
showing a divergently large value also in the strong field limit 
(see also Fig.~\ref{fig:Br}). 
Therefore, even at a small photon energy, 
the dielectric constant approaches the limiting value $\epslim$ 
when the magnetic field is strong enough.

In the limit of zero frequency $\ome^2 = 0$, 
both $r_\perp^2$ and $r_\parallel^2$ are vanishing, and the scalar 
coefficient function $\chi_1^{\rm LLL}$ has a simple 
form (see Eq.~(\ref{eq:chi1_LLL})), 
\begin{eqnarray}
\chi_1^{\rm LLL}(0) = \frac{ \alpha }{ 3 \pi } \Br
\ \ ,
\end{eqnarray}
which indicates a monotonic increase with respect to $\Br$. 
Thus, $\chi_1^{\rm LLL}$ becomes of the order $\chi_1^{\rm LLL} \sim1$ 
if the magnetic field is as strong as $\Br \sim 3 \pi / \alpha \sim 10^3$. 
Beyond this field strength, 
$\chi_1$ starts dominating unities in the denominator and numerator in Eq.~(\ref{eq:eps_LLL}). 
Then, the dielectric constant and thus the refractive index 
would start approaching the limiting values $\epslim $ and $\nlim$, respectively. 
The upper panel in Fig.~\ref{fig:ref_Br} shows 
sizable modification of the refractive index already 
around $\Br \sim 10^3$ at zero and low energies 
($\Br = 3 \pi / \alpha$ gives $n_\parallel = 4/3$ when $\theta=\pi/4$).

As the photon energy becomes large beyond 
$\ome ^2 = 1/( 1- \cos^2\theta)$ ($\ome^2 = 2$ when $\theta=\pi/4$ 
in Fig.~\ref{fig:ref_Br}), 
the refractive index is subject to significant modification 
even in a relatively weak regime, $\Br \alt 100$. 
This is because the squared momentum approaches the threshold 
$\rp (\epara) = 1$, 
and thus $\chi_1$ increases divergently (see Fig.~\ref{fig:chi1_LLL}).


Lower panel in Fig.~\ref{fig:ref_Br} shows the real part of the refractive index 
on the unstable branch. 
In this plot, each line has an end point 
at a certain $\Br$, 
because the unstable branch does not exist below a connection point 
between the stable and unstable branches, 
as shown in Fig.~\ref{fig:ref500}. 
Black long-dashed curve 
shows location of the connection point 
specified by a pair of photon energy and magnetic field strength. 
As found in a comparison between Fig.~\ref{fig:full500} and Fig.~\ref{fig:full5000}, 
the connection point shifts to a larger photon energy as $\Br$ increases. 
Therefore, for a given photon energy, 
unstable branch does not exist above a certain magnetic field strength. 
Although peaks apparently look sharper as the photon energy increases, 
this is due to a logarithmic scale for the horizontal axis. 


\begin{figure}[t]
	\begin{center}
	\includegraphics[width=0.7\hsize]{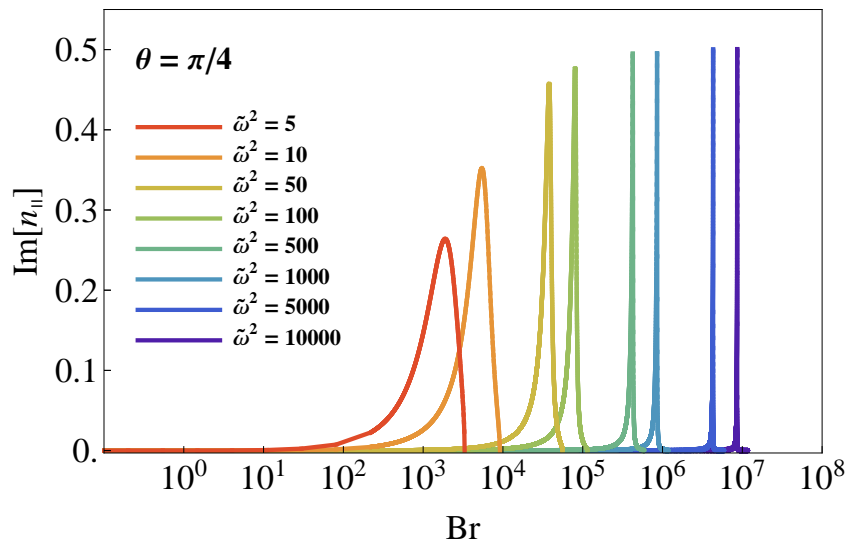}
	\end{center}
	  \vspace{-0.5cm}
\caption{
$\Br$ dependence of the imaginary part of $n_\parallel$ 
on the unstable branch. Curves show those at $\theta = \pi/4$ 
and various photon energies. 
}
\label{fig:ref_Br_im}
\end{figure}


\subsubsection{Imaginary part}

Figure~\ref{fig:ref_Br_im} shows $\Br$-dependence of the imaginary part 
of the refractive index on the unstable branch. 
As in the case of the real part shown in the lower panel in Fig.~\ref{fig:ref_Br}, 
there is a peak structure just below the end point 
of each curve, reflecting a prominent photon decay rate near the connection point. 
As we found in the comparison between Fig.~\ref{fig:full500} and Fig.~\ref{fig:full5000}, 
the threshold obtained in the self-consistent solution (the connection point) 
shifts more distantly from the naive one $\tilde \omega^2_{\rm th} = 1/\sin^2\theta$ 
to a large photon energy when the magnetic-field strength is larger. 
Consistently to what we found there, 
a stronger magnetic field gives rise to the peak for an energetic photon in Fig.~\ref{fig:ref_Br}.


\begin{figure}[t]
  \begin{center}
   \includegraphics[width=0.7\hsize]{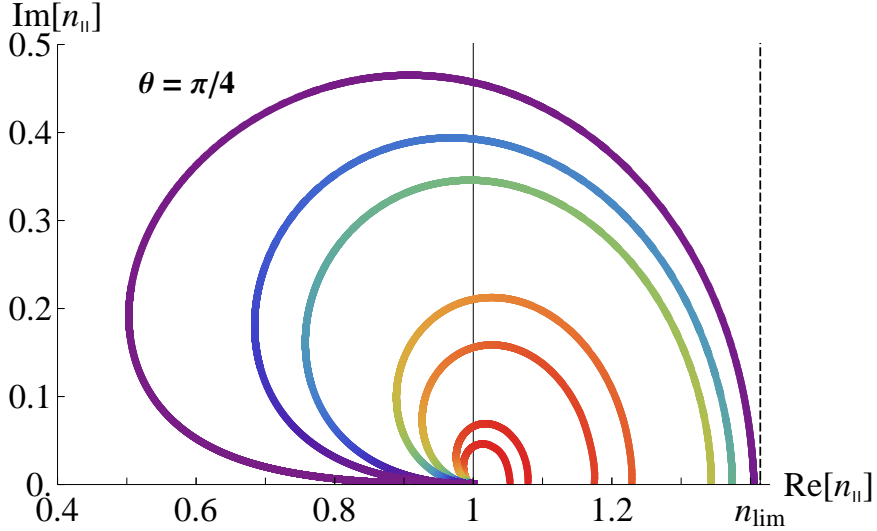}
  \end{center}
  \vspace{-0.3cm}
\caption{
Trajectories in a complex $n_\parallel$-plane showing its magnitudes on the unstable branch: 
The propagation angles are commonly fixed at $\theta = \pi/4$, and 
the magnetic field strength on each trajectory is taken as $\Br = \{ 50, 100, 500, 1000, 5000, 10000, 50000 \}$ 
from the innermost to the outermost. 
On each curve, the photon energy increases from the right to left edges, 
where the right edge corresponds to the connection point between the stable and unstable branches. 
}
 \label{fig:ref_ri}
\end{figure}


\subsubsection{Real vs imaginary part on the unstable branch}

Figure~\ref{fig:ref_ri} shows a relation between real and imaginary parts 
of the complex refractive index on the unstable branch. 
The magnitude of the external magnetic field is fixed on each arc, 
and a radius increases as the magnetic field strength becomes larger. 
A gradation of colors transits from red to violet as the photon energy increases, 
of which scale is fixed so that the full range from red to violet appears on the innermost arc. 
Every arc has a left edge located on $\epara = 1 + 0 \cdot i$, 
because the refractive index on the unstable branch converges to unity 
in case of the strong-field limit within the LLL approximation, 
when the photon energy becomes large beyond the first threshold 
and stays small enough not to approach the second threshold. 
The opposite edge corresponds to the connection point between the stable and unstable branches. 
The imaginary part of the refractive index vanishes at this point, 
and the real part becomes large as the field strength becomes large. 
Magnitudes of the modification of the real and imaginary parts 
are comparable to each other at a fixed $\Br$.

\subsection{Decay length}


\begin{figure}[t]
  \begin{center}
   \includegraphics[width=0.65\hsize]{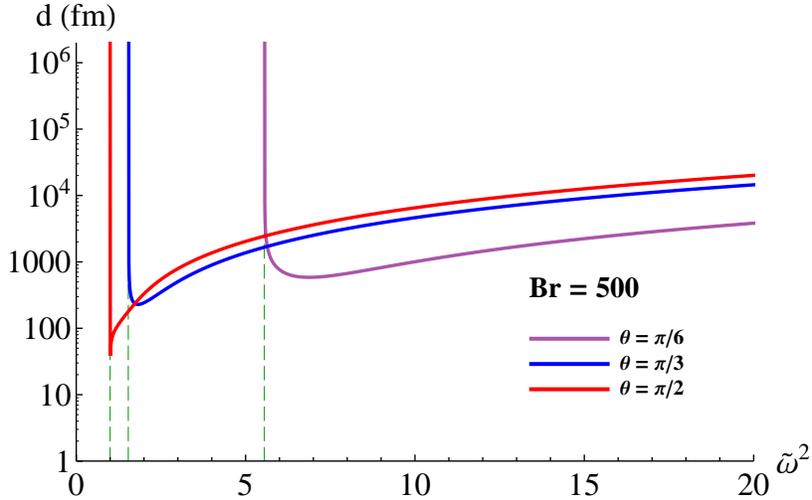}
	\end{center}
\caption{Decay length of photon propagation in strong magnetic field: 
Photon decays within order of 0.1 -- 10 picometer above the threshold, 
while magnetar has magnetic field extending in a macroscopic scale.}
\label{fig:length}
\end{figure}

Lastly, let us show the decay length defined in Eq.~(\ref{eq:decay_length}). 
Figure~\ref{fig:length} shows the decay length at a magnetic field 
strength $\Br = 500$ and propagation angles, $\theta  = \pi/6, \pi/3 , \pi/2$. 
Photon energy at the decay threshold becomes small as the angle increases, 
because of an angle dependence of the threshold condition, $\rp = 1$. 
While ``magnetar" is thought to be accompanied by 
a strong magnetic field\footnote{
Here, physical scale is provided by the critical magnetic field strength 
defined by electron mass. 
} 
of order $\Br \sim 100$ extending in a macroscopic scale, 
plotted lines show that photon decays within order of 0.1 -- 10 picometer 
if photon energy is above the threshold. 
This indicates a drastic modification of gamma-ray and dilepton spectra emitted from magnetars. 

\section{Summary and discussions}



We investigated the vacuum birefringence phenomena 
in strong magnetic fields on the basis of the analytic representation 
of the vacuum polarization tensor obtained in paper I \cite{HI1}. 
In particular, we analyzed in detail the phenomena in the 
lowest Landau level (LLL) approximation where both the dielectric 
constant ($\epara$) and the refractive index ($n_\parallel$) of 
a propagating photon in the mode parallel to the external magnetic 
field deviate from unity and become complex. The LLL contribution 
corresponds to the first term of the polarization tensor which 
is represented as the infinite sum with respect to the Landau 
levels, and turned out to be a very good approximation when the 
magnetic field is strong enough $B/B_c\simge 1$ with $B_c$ being 
the critical magnetic field strength for the relevant fermions.

In the LLL approximation, there is a threshold in the 
photon energy beyond 
which both $\epara$ and $n_\parallel$ acquire imaginary parts
and thus the decay of a real photon into a pair of a fermion and an 
antifermion (in the LLLs) becomes possible. With the explicit 
analytic expression for the polarization tensor, we were able to 
thoroughly 
inspect properties of $\epara$ and $n_\parallel$ such as the 
dependences on the propagating angle and the magnetic-field strength,
and also how the imaginary parts appear. 

We found that the equations that define $\epsilon$ (or equivalently 
$n$) with respect to the scalar coefficient functions in the polarization 
tensor are implicit functions
of $\epsilon$ (or $n$), and must be solved self-consistently in terms of 
$\epsilon$ (or $n$). This self-consistent procedure is 
indispensable for accurate description of $\epsilon$ and $n$ when 
the deviation of $\epsilon$ and $n$ from unity is 
large, which is 
realized in the limit of strong fields or high-energy photons. 
Physically, this procedure is necessary for consistently taking into account 
the back reaction from the distorted Dirac sea 
in response to an incident photon field. 
Namely, an induced polarization and the photon decay bring in modifications of 
the external momentum, and thus dispersion, of the incident photon.

\begin{figure}[t]
  \begin{center}\hspace*{-5mm}
   \includegraphics[width=0.47\hsize]{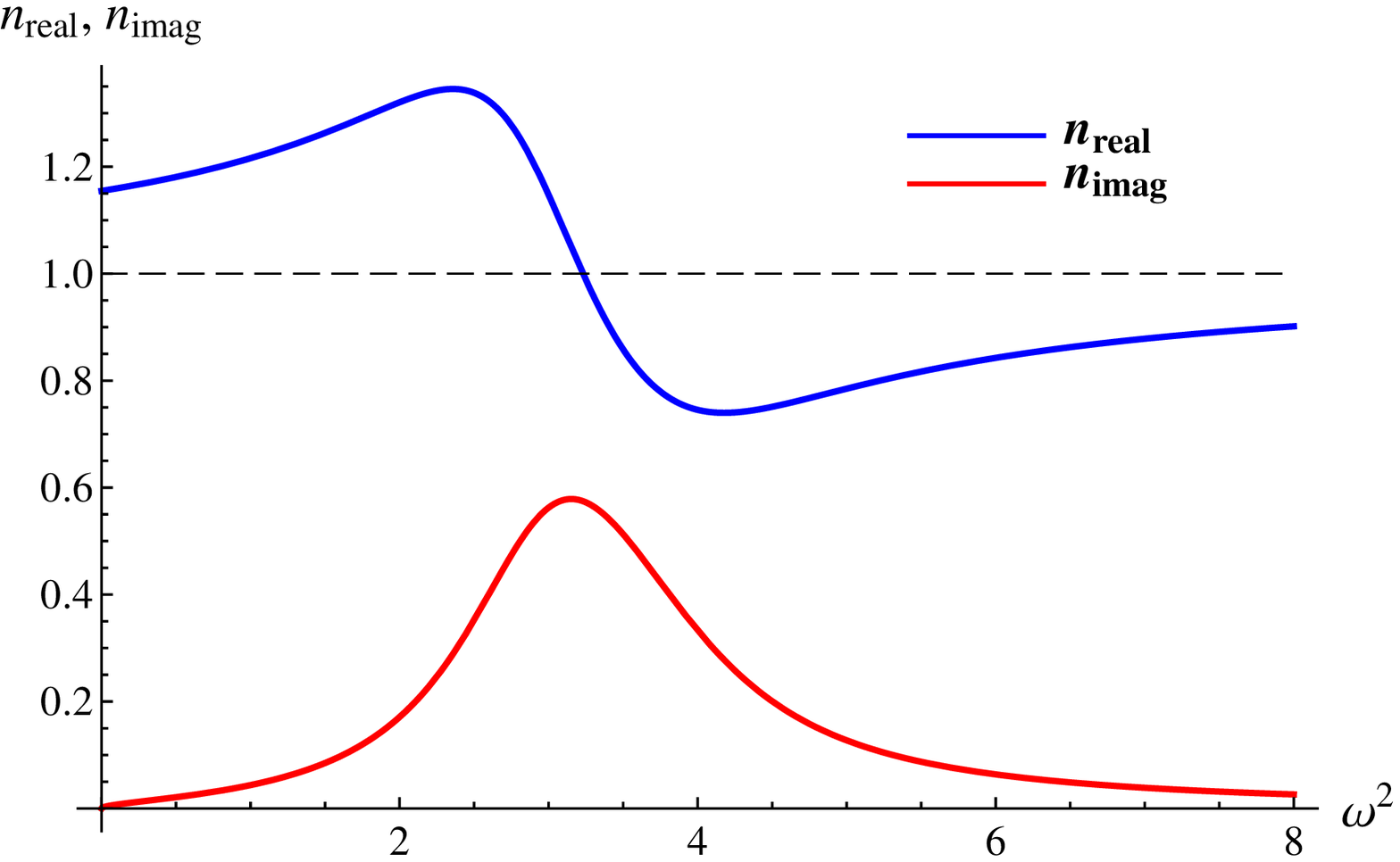}\quad 
   \includegraphics[width=0.47\hsize]{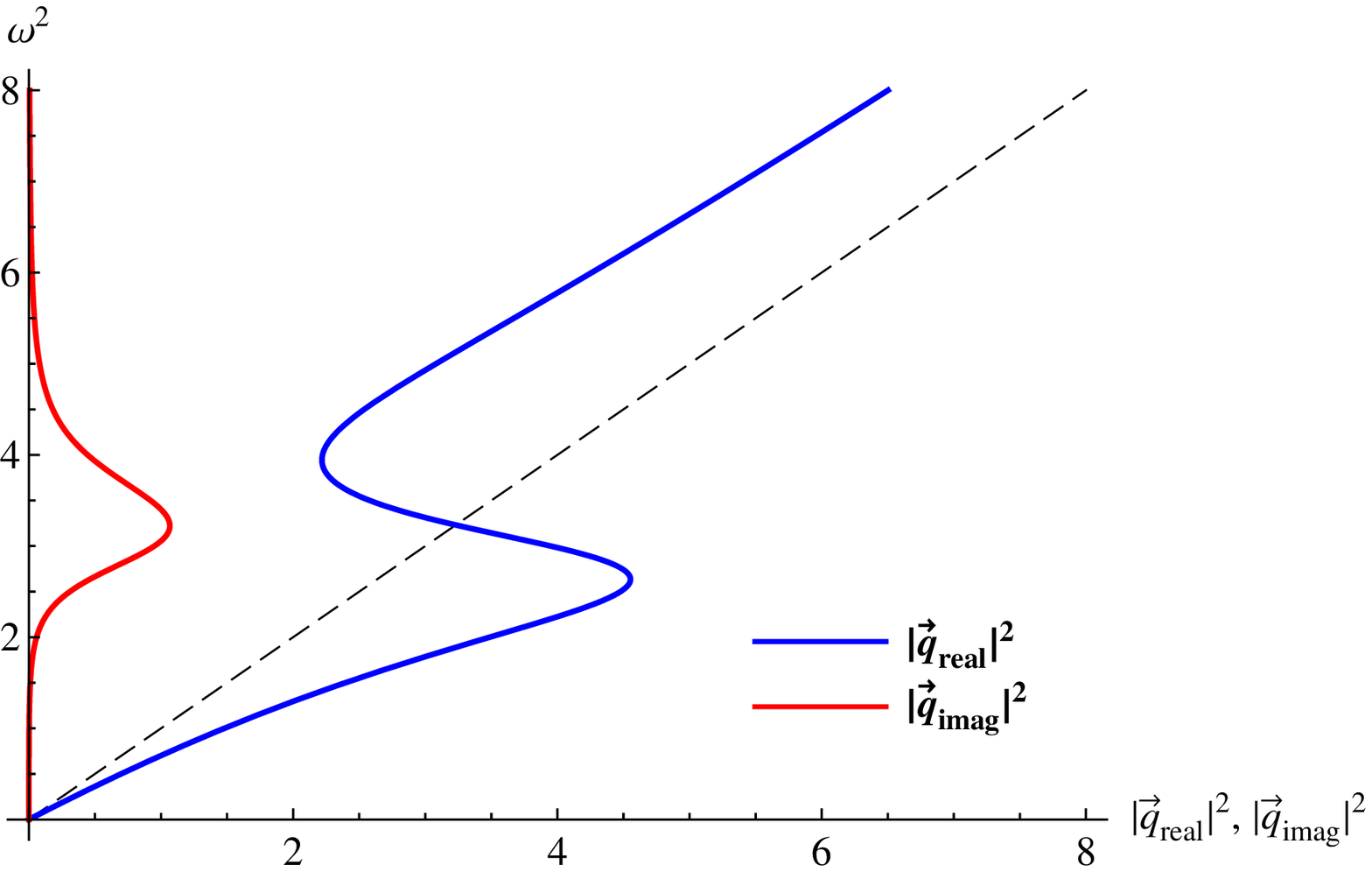}
	\end{center}
\caption{
(Left) Refractive indices of a simple damped harmonic oscillator model. 
(Right) Dispersion of a propagating photon in a dielectric medium. Blue and red lines are, respectively, the real and imaginary parts of the refractive index. Dashed lines are the refractive index $n=1$ (left) and the dispersion $\omega=|q_{\rm real}|$ of a photon in a vacuum (right). 
Energy and momentum are taken in arbitrary unit for an illustration. 
}
  \label{fig:oscillators}
\end{figure}

\begin{figure}[t]
  \begin{center}
   \includegraphics[width=0.65\hsize]{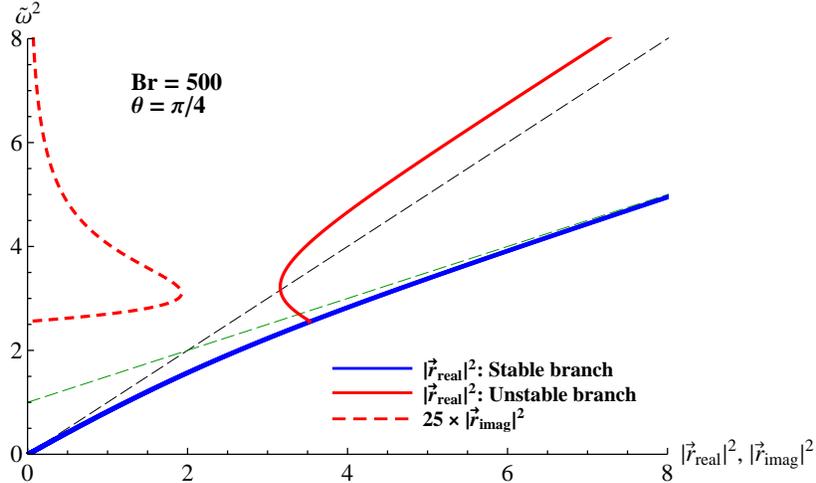}
	\end{center}
\caption{
Dispersion relation of a photon propagating in a strong magnetic field. 
The propagation angle and the magnetic field strength, shown in the legend, 
are taken as in Fig.~\ref{fig:ref500}. }
  \label{fig:disp2}
\end{figure}

It is quite instructive to recognize that emergence of complex refractive indices 
can be seen even in a simple example, the dipole oscillator model, of an optical medium 
(see Refs.~\cite{Hecht,Fox} for pedagogical descriptions). 
The simplest case assumes that the medium is composed of many damped harmonic oscillators 
with a single resonant frequency $\omega_0$. 
The refractive index of this model has both real and imaginary parts, 
as shown in the left panel of Fig.~\ref{fig:oscillators}. 
The imaginary part appears when the real part rapidly varies around the resonant frequency $\omega=\omega_0$, 
which corresponds to a negative slope in the dispersion (see the right panel of Fig.~\ref{fig:oscillators}). 
This rapid change of the real part across unity 
and emergence of a peak of the imaginary part are qualitatively similar to the behaviors seen in Fig.~\ref{fig:ref500}. 
Consequently, 
a dispersion curve in Fig.~\ref{fig:disp2}, following the blue solid curve from the low photon energy 
up to the junction point and then the red curve, 
corresponds to the blue line in the right panel of Fig.~\ref{fig:oscillators}. 
The dispersions in our result and the dipole oscillator model 
have a similar structure to each other, except two differences: that is, 
there is a definite threshold for the emergence of the imaginary part located on the lowest Landau level 
and the ``stable" branch extends into infinite $\tilde \omega^2$ giving a double-valued refractive index 
above the threshold.

Now let us suppose that a photon enters 
into the dielectric medium from the ordinary vacuum, 
as an analogue of the case incident into strong magnetic fields. 
Dispersion relation of the photon will change from the linear relation 
(a dashed line in the right panel of Fig.~\ref{fig:oscillators}) to 
that in the medium (a blue curve in Fig.~\ref{fig:oscillators}). 
At the boundary, according to the Huygens-Fresnel principle, the direction of 
the light front will be modified consistently to the change of light velocity, 
while the frequency (photon energy) is conserved. Therefore, the incoming 
photon with $(\omega_{\rm vac},q_{\rm vac})$ on the linear dispersion will 
move to $(\omega_{\rm med}=\omega_{\rm vac}, q_{\rm med})$ on the curved 
dispersion line ($q_{\rm med}\neq q_{\rm vac}$). 
This argument does not prohibit transition to any in-medium dispersion as long as the energy conservation is satisfied, 
and, if the incident light is intense enough, 
a nonlinear response of the dielectric medium to the light generates higher harmonic waves, 
so that even transitions to other frequencies could become possible. 
Then, what happens in the case 
with the vacuum birefringence in strong magnetic fields? 
Having performed self-consistent treatment, we found the refractive index as a double-valued function of 
the photon energy that is composed of stable and unstable branches shown in 
Fig.~\ref{fig:ref500} or in Fig.~\ref{fig:disp2}. 
If one considers an incoming photon from the vacuum into the region with a strong magnetic field, 
one encounters a problem that there are two possible transitions 
(one is onto the red curve and the other, blue) when the photon energy is high enough. 
Assuming that 
the transitions to both the branches are possible, it would be natural 
that the transition to the closer branch is 
preferred. 
Thus, which branch is realized depends on 
the dispersion of the incoming photon outside the magnetic field 
as well as the detailed structure of the dispersion in the magnetic field. 
As far as the photon enters 
from the vacuum, 
we expect that the unstable branch would be more preferably 
realized in case shown in Fig.~\ref{fig:disp2}.

We also call attention to a similarity between dispersions in our result and of ``polariton" 
which is a resonant state arising in a coupling between a quasi-particle and a photon. 
Dispersion relation of a polariton typically has two separate branches with a level repulsion 
at the crossing point of the original dispersions of those constituent particles. 
More specifically, there are some well-known polaritons, e.g., 
called ``exciton polariton", ``surface plasmon polariton" and 
``phonon polariton", excited when a photon couples to a particle-hole excitation in semiconductors, 
a quantized plasma oscillation in the vicinity of metal surfaces 
and a phonon in crystals, respectively (see Sec.~4, 7.5 and 10.3 in Ref.~\cite{Fox}). 
They have been indeed observed in experiments. 
However, polaritons can be excited only when the dispersion of a probe photon incident into the substances 
has an intersection with the polariton dispersion curves, 
while we argued on the basis of an analogy with the classical dipole oscillator 
that the transition to any branch is in principle possible within a certain probability. 
Nevertheless, both of these analogies 
indicate that the dispersion outside the medium also plays an important role 
as an initial condition for realization of the in-medium propagating modes.

Although these analogies seem not to be totally suitable to the present case, 
they could provide clues to grasp intuitive understandings and prospects. 
A ``selection rule" for the transitions onto the branches will be studied elsewhere.

Lastly, let us compare our results with the refractive indices which 
we encounter in ordinary life, and discuss possible applications to the 
physical situations accompanied with strong magnetic fields. 
The refractive index acquires 
a large value in the strong field limit. 
For example, at the angle $\theta=\pi/4$, the refractive index 
$n_\parallel$ keeps increasing with stronger magnetic fields, and 
approaches the limiting value 
$n_{\rm lim}=\sqrt2$. This is comparable to the values which we encounter 
in ordinary life. To name a few, atmosphere of the earth 
($1$ atm, $0\ {}^\circ\mathrm{C}$) and water ($20\ {}^\circ\mathrm{C}$) 
have refractive indices, 
$n_{\rm air} = 1.000293$ and $n_{\rm water} = 1.333$, 
respectively, and ``calcite" known as a representative birefringent 
material has refractive indices 
$n_{\rm o}  = 1.6584$ and $n_{\rm e}  = 1.4864$ 
for ordinary and extraordinary modes, respectively. 
At a magnetic field strength $\Br \sim (m_\pi/m_e)^2 \sim 10^{4-5} $ 
which could be realized in the ultrarelativistic heavy-ion collisions\footnote{
$m_\pi$ and $m_e$ are masses of a pion $m_\pi \sim 140$ MeV and 
an electron $m_e \sim 0.5$ MeV, respectively.} 
\cite{KMW,Sko,DH,Itakura_PIF}, 
the refractive index is close to the limiting value $n_{\rm lim}=\sqrt2$, 
and thus can be larger than that of gas, and even comparable to those of 
liquid and solid.

An accurate description of the complex refractive index across the lowest 
threshold will be directly applicable to the studies of intriguing 
phenomena in magnetars whose magnetic field is estimated to be two orders 
larger than the critical magnetic field strength \cite{TD,MagnetarReview}. 
In particular, descriptions of both the real and imaginary parts 
will work as robust building blocks necessary for investigating 
the interplay and competition among photon splitting, 
vacuum birefringence and photon decay \cite{Adl70,Bar}.

Besides, as mentioned above, 
the ever strongest magnetic field will be created 
in the ultrarelativistic heavy-ion collisions in RHIC and LHC 
experiments. 
Since the magnitude of the magnetic field could be four orders larger than 
the critical magnetic field, radiations from the created matter in collision 
events would interact with the extremely strong magnetic 
fields, and bring out the information of the early-time dynamics 
\cite{Tu,IHhbt}. However, photons created in heavy-ion collisions 
have, in general, large energies. Thus we need to extend 
the present work towards 
including important contributions among all the Landau levels shown 
in paper I. 

In application to laser physics \cite{TopicalReviewLASER}, 
it will be very important to include effects of a strong electric field 
as well as a magnetic field. 
For this sake, basic concepts and techniques examined in this series of 
papers could be useful to find differences out of similarities between 
expressions 
of the vacuum polarization tensors in the presence of external electric 
and magnetic fields.


\section*{Acknowledgements}
The research of KH is supported by the Korean Ministry of Education through the BK21 Program. 
This work was also partially supported by Korea national research foundation under 
grant number KRF-2011-0030621 
and ``The Center for the Promotion of Integrated Sciences (CPIS)" of Sokendai. 

\appendix

\section{Dependence of the dielectric constant on $\Br$ and $\theta$}

\label{sec:eps_Br}


\begin{figure}[t]
 \begin{minipage}[t]{0.47\hsize}
  \begin{center}
   \includegraphics[width=\hsize]{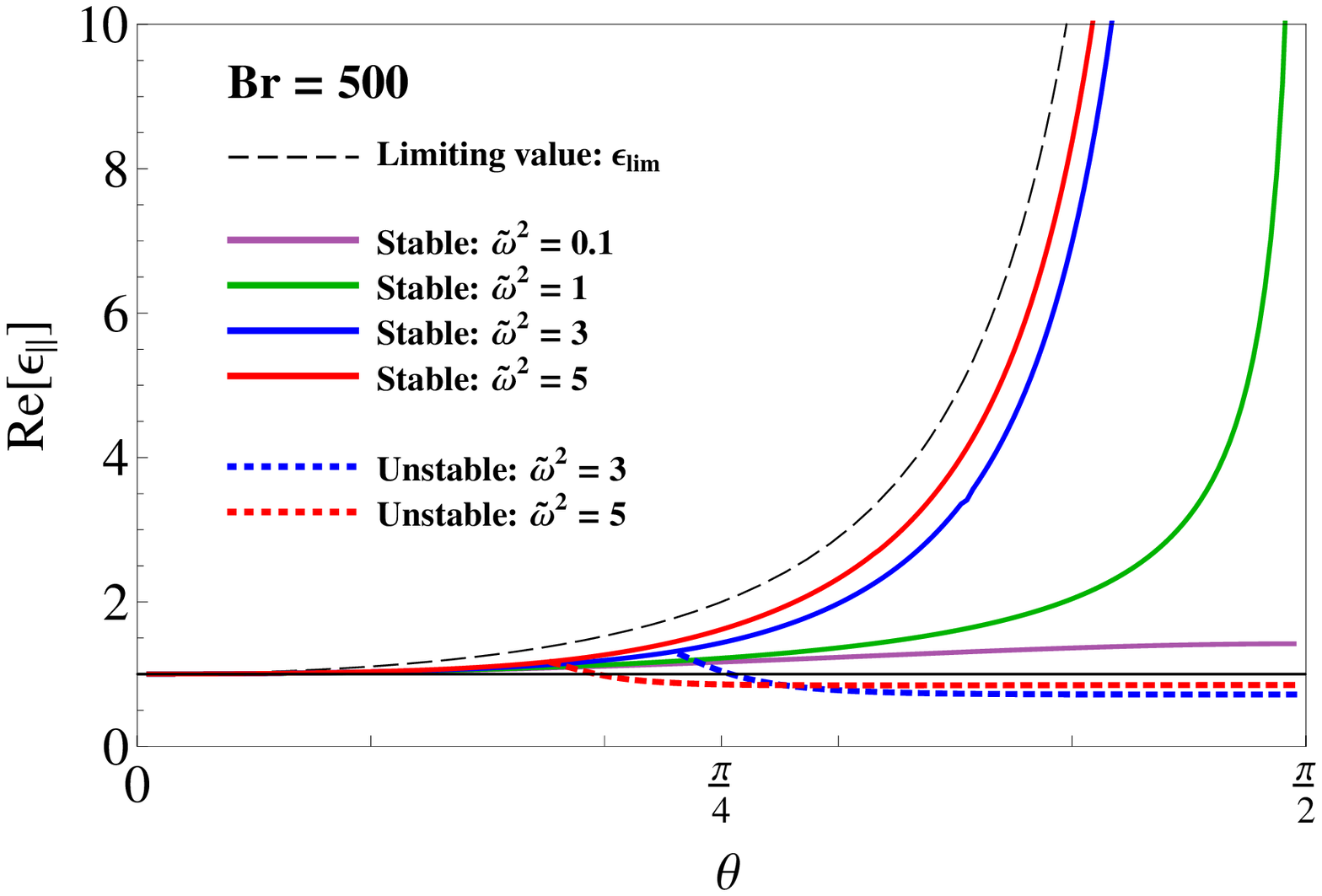}
  \end{center}
 \end{minipage}
  \hspace{0.02\hsize}
	\begin{minipage}[t]{0.47\hsize}
		\begin{center}
			\includegraphics[width=\hsize]{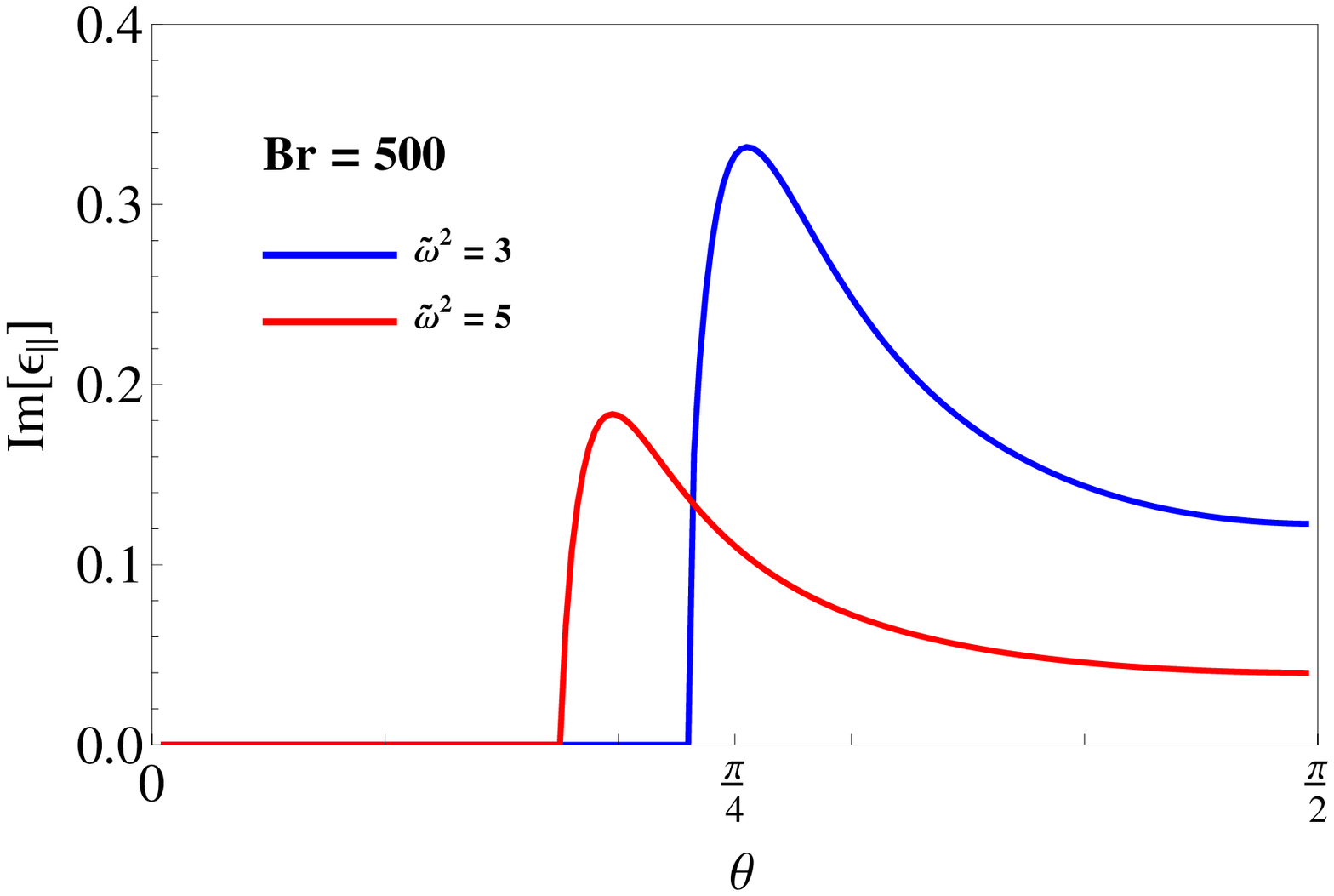}
		\end{center}
	\end{minipage}  
\vspace{-4mm}
\caption{Dependence of $\epara$ on photon's 
propagation angle $\theta$: 
Parameters are the same as in Fig.~\ref{fig:angle}. 
Dashed line shows the limiting value (\ref{eq:eps_limit}). 
}
\label{fig:eps_angle}
\end{figure}



\begin{figure}[t]
 \begin{minipage}[t]{0.49\hsize}
  \begin{center}
   \includegraphics[width=\hsize]{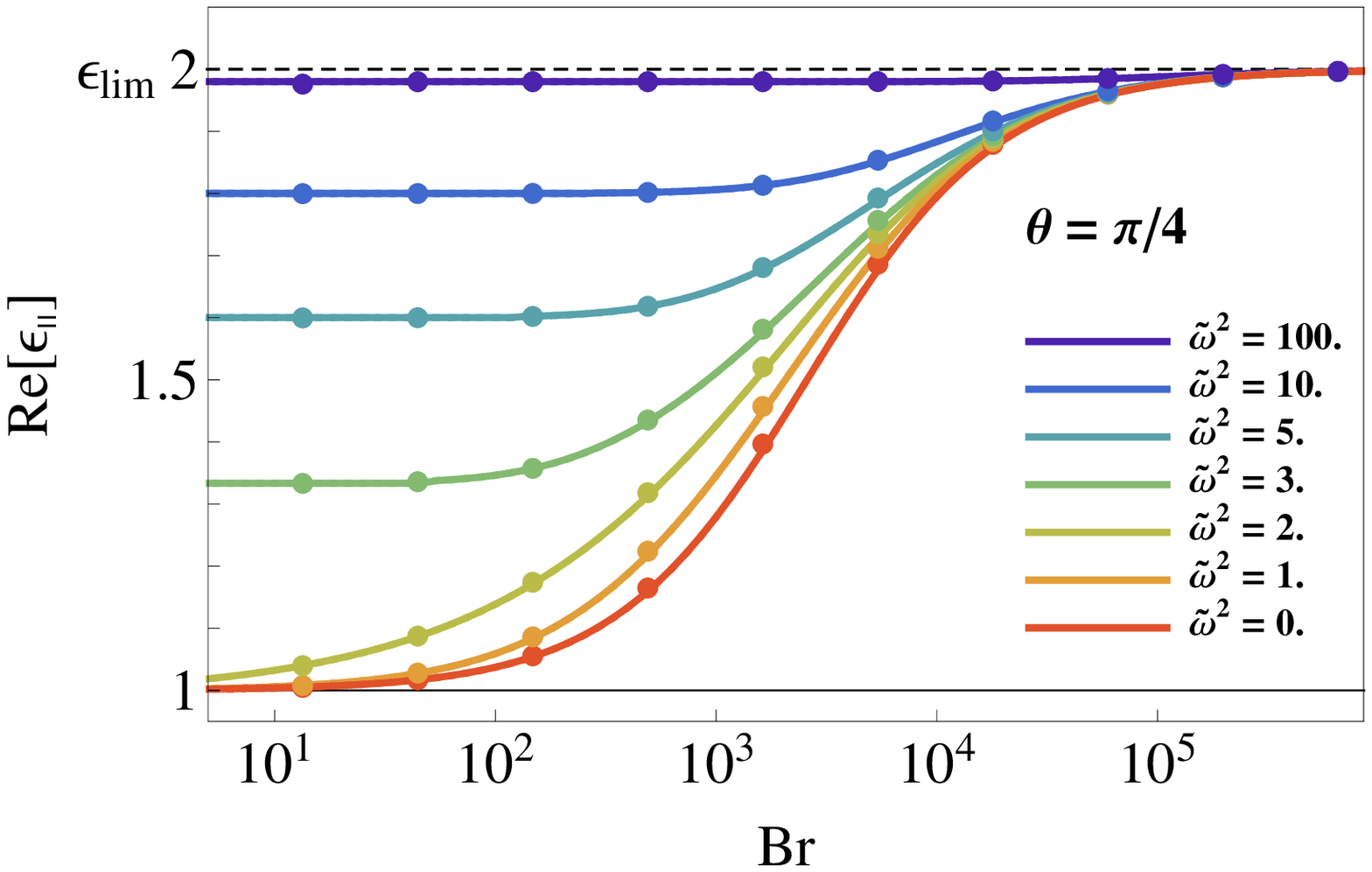}
  \end{center}
 \end{minipage}
%
 \begin{minipage}[t]{0.49\hsize}
	\begin{center}
	\includegraphics[width=\hsize]{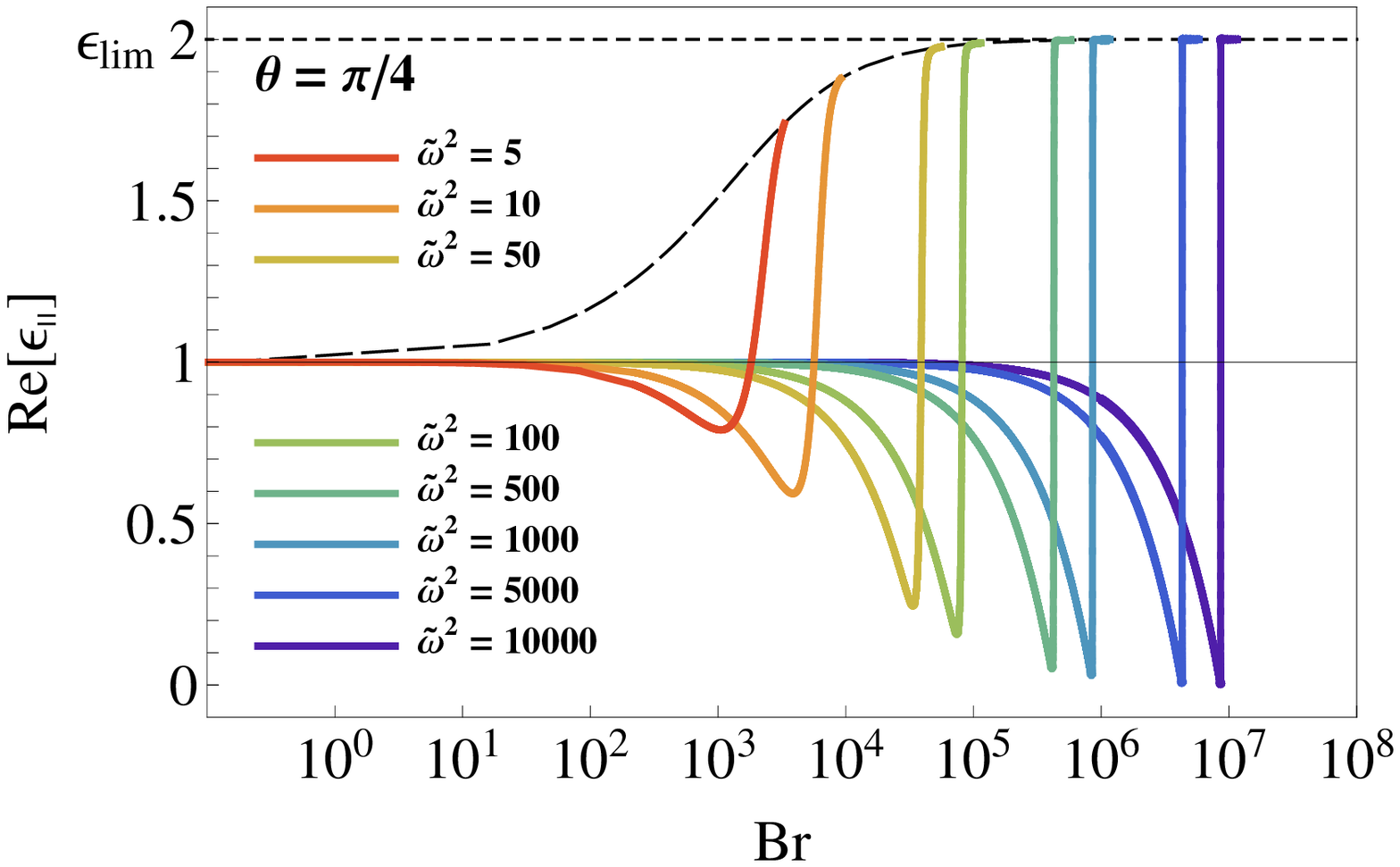}
	\end{center}
 \end{minipage}  
\vspace{-4mm}
\caption{$\Br$ dependence of the real part of 
$\epara$ on the stable (left) and unstable (right) branches. 
Photon's propagation angle is fixed at $\theta = \pi/4$, 
and photon energies are varied as indicated in the panels. 
Dielectric constant on the stable branch, obtained from the 
numerical integration, is shown with dots in the left panel. 
}
\label{fig:eps_Br}
\end{figure}

\begin{figure}[t]
 \begin{minipage}[t]{0.48\hsize}
	\begin{center}
	\includegraphics[width=\hsize]{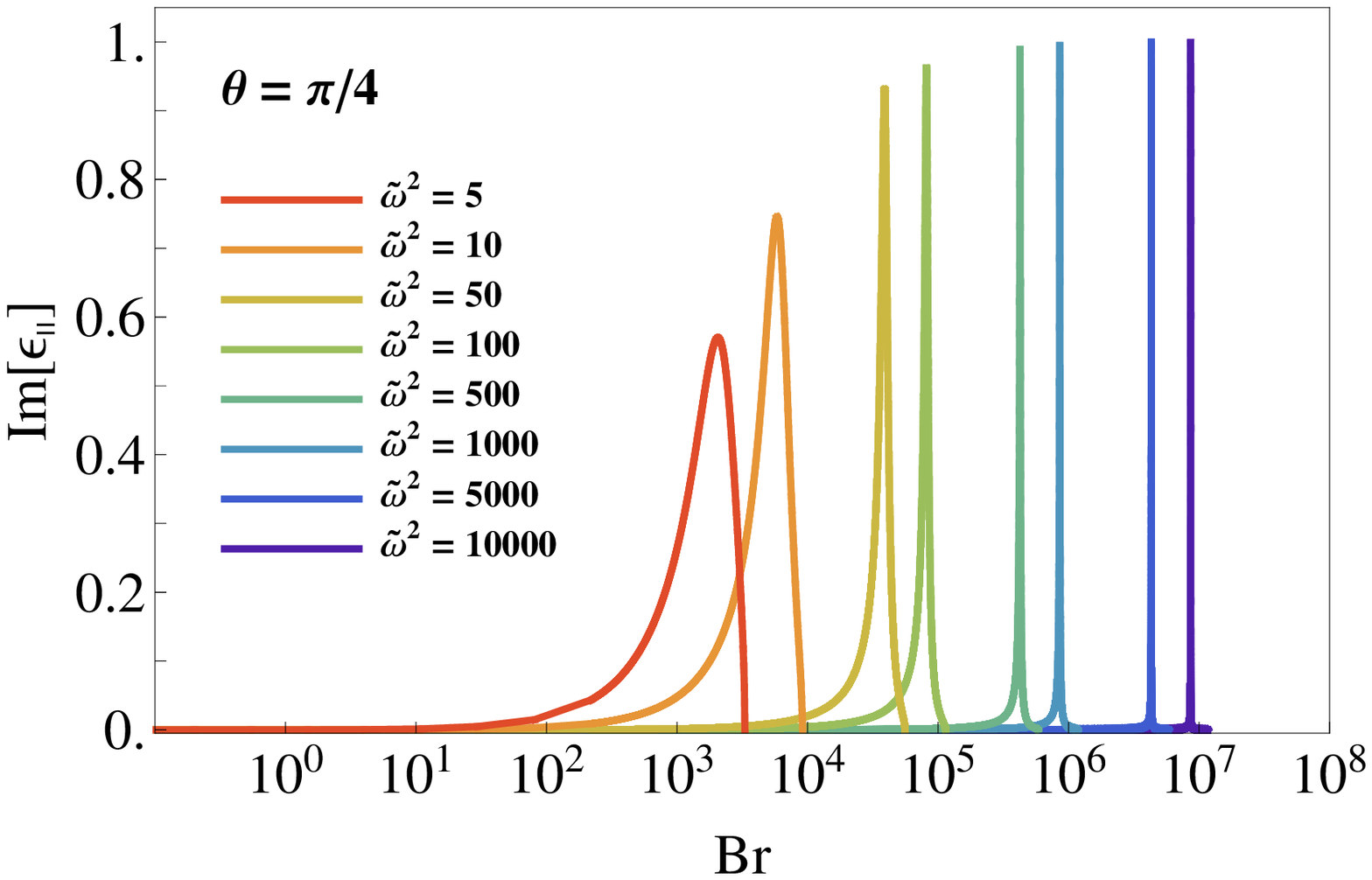}
	\end{center}
\vspace{-6mm}
\caption{$\Br$-dependence of the imaginary part of $\epara$ 
on the unstable branch. Parameters are the same as in 
the right panel of Fig.~\ref{fig:eps_Br}. 
}
\label{fig:eps_Br_im}
\end{minipage}
\hspace{0.2cm}
\begin{minipage}[t]{0.48\hsize}
  \begin{center}
   \includegraphics[width=\hsize]{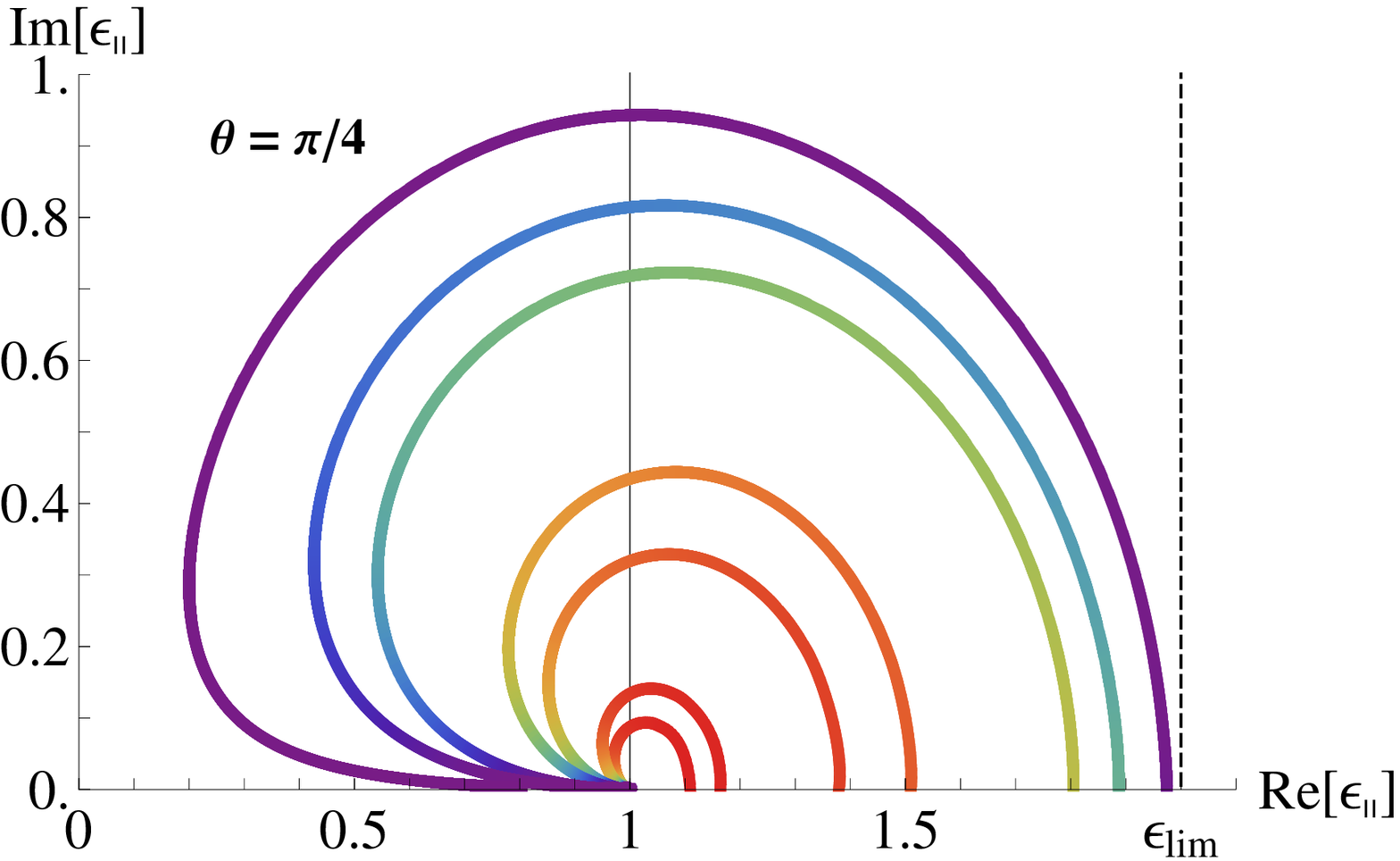}
  \end{center}
\vspace{-6mm}
\caption{
Trajectories on a complex $\epsilon_\parallel$-plane. 
Parameters are the same as in Fig.~\ref{fig:ref_ri}. }
\label{fig:eps_ri}
 \end{minipage}
\end{figure}


In this supplementary section, 
we show dependences of the dielectric constant $\epara$ on 
the propagation angle $\theta$ and the magnetic-field strength $\Br$. 
By self-consistently solving Eq.~(\ref{eq:eps_LLL}), we obtained 
plots shown in Figs.~\ref{fig:eps_angle} -- \ref{fig:eps_ri}, 
which are then mapped to the refractive index thought the 
relations (\ref{eq:ref_real}) and (\ref{eq:ref_imag}) to yield 
Figs.~\ref{fig:angle} -- \ref{fig:ref_ri}. 
Parameters in these plots are taken as the same values as 
those in the corresponding plots of the refractive index in 
Fig.~\ref{fig:angle} -- \ref{fig:ref_ri}. 
The dependence on the photon's propagation angle is shown in 
Fig.~\ref{fig:eps_angle}. 
The real part of the dielectric constant on the stable and unstable 
branches are shown in the left and right panels in Fig.~\ref{fig:eps_Br}, 
respectively, while the imaginary part on the unstable branch is 
shown in Fig.~\ref{fig:eps_Br_im}. 
Magnitudes of the real and imaginary parts on the unstable 
branch are shown together in Fi.g~\ref{fig:eps_ri} as contours 
in the complex $\epara$-plane.


\section{Remark on other Landau levels}

We remark on structures near the thresholds in other Landau levels which 
have the same property as that of the lowest Landau level. 
In the infinite series expression of the scalar coefficient 
functions $\chi_i$ with all the Landau levels 
(see Sec.~IV 
in paper I \cite{HI1}), 
we find a relation between the coefficient functions $\chi_0 = - \chi_2$, 
if we pick up 
an individual contribution 
specified by $\ell = 0$ and $n (\neq 0)$. 
It immediately follows from Eq.~(\ref{epsilon_perp}) that 
the dielectric constant of the perpendicular mode is then given by 
\begin{eqnarray}
\epsilon_\perp = \frac{ 1 + \chi_0 }{ 1 + \chi_0 \cos^2 \theta }
\ \ .
\end{eqnarray}
The right-hand side of the above expression has the same form as Eq.~(\ref{eq:eps_LLL}), 
except that the scalar function $\chi_1$ is replaced by $\chi_0$. 
Because $\chi_0$ diverges at these thresholds as $\chi_1$ does 
at the lowest threshold, 
the dielectric constant has a liming value, 
\begin{eqnarray}
\lim_{\rp \rightarrow s_+^{0 n} } \epsilon_\perp (\rp) = \frac{ 1 }{ \cos^2 \theta }
\ \ ,
\label{eq:eps_lim_perp}
\end{eqnarray}
when the photon momentum approaches the thresholds 
as $\rp \rightarrow ( 1 + \sqrt{1 + 2n \Br } )^2 /4 \equiv  s_+^{0 n}  $. 
The right-hand side of Eq.~(\ref{eq:eps_lim_perp}) is exactly 
the same as that of Eq.~(\ref{eq:eps_limit}). 
Therefore, the dielectric constant $\eperp$ in the perpendicular mode 
would, in the region around the thresholds at $\ell = 0$ and $n ( \neq 0)$, 
have the same structure in the photon energy dependence as that of 
the longitudinal mode $\epara$ around the lowest threshold.




\input{bib}
\end{document}

%% file: head.tex
\author{Koichi Hattori}\email{\tt khattori@yonsei.ac.kr}

\affiliation{
Institute of Physics and Applied Physics, 
Yonsei University, Seoul 120-749, Korea 
}

\author{Kazunori Itakura }\email{\tt kazunori.itakura@kek.jp}
\affiliation{Theory Center, IPNS, 
High energy accelerator research organization (KEK), 
1-1 Oho, Tsukuba, Ibaraki 305-0801, Japan 
and \\
Department of Particle and Nuclear Studies, 
Graduate University for Advanced Studies (SOKENDAI), 
1-1 Oho, Tsukuba, Ibaraki 305-0801, Japan}


\vspace*{5mm}
\title{Vacuum birefringence in strong magnetic fields: \\
(II) Complex refractive index from the lowest Landau level}


\date{\today}

\begin{abstract}
We compute the refractive indices of a photon propagating in strong 
magnetic fields on the basis of the analytic representation of the 
vacuum polarization tensor obtained in our previous paper. When 
the external magnetic field is strong enough for the 
fermion one-loop diagram of the 
polarization tensor to be approximated by the lowest Landau level, 
the propagating mode in parallel 
to the magnetic field is subject to modification: The refractive 
index deviates from unity and can be very large, 
and when the photon energy is large enough, the refractive index 
acquires an imaginary part indicating decay of a photon into a 
fermion-antifermion pair. We study dependences of the refractive 
index on the propagating angle and the magnetic-field strength. 
It is also emphasized that a self-consistent treatment of the 
equation which defines the refractive 
index is indispensable for accurate description of the refractive index.
This self-consistent treatment physically corresponds to consistently 
including the effects of back reactions of the 
distorted Dirac sea in response to the incident photon.

\end{abstract}


\preprint{KEK-TH-1588}



\maketitle

%% file: main.bbl
\begin{thebibliography}{99}


\bibitem{HI1} K. Hattori and K. Itakura, ``{\it Vacuum birefringence 
in strong magnetic fields: (I) Photon polarization tensor with all the Landau levels}",  
Ann. Phys. {\bf 330} (2013) 23-54. {\tt arXiv:\,1209.2663\,[hep-ph].}


\bibitem{MS76} D. Melrose and R. Stoneham, Nuovo Cimento A {\bf 32} (1976) 435.


\bibitem{Tsa} W. Tsai, Phys. Rev. D {\bf 10} (1974) 2699. 
\bibitem{Urr78} L. F. Urrutia, Phys. Rev. D {\bf 17} (1978) 1977. 

\bibitem{Adl1} S. L. Adler, Ann. Phys. {\bf 67} (1971) 599-647. 

\bibitem{TE} W. Tsai and T. Erber, Phys. Rev. D {\bf 10} (1974) 492; {\it ibid}., Phys. Rev. D {\bf 12} (1975) 1132. 

\bibitem{DR} W. Dittrich and M. Reuter, Lect. Notes Phys. {\bf 220} (1985) 1-244. 
\bibitem{DG} W. Dittrich and H. Gies, Springer Tracts Mod. Phys. {\bf 166} (2000) 1-241. 

\bibitem{Sch} J. S. Schwinger, Phys. Rev. {\bf 82} (1951) 664-679. 

\bibitem{Fuk} K. Fukushima, Phys. Rev. D {\bf 83} (2011) 111501. 

\bibitem{GMS96} V. P. Gusynin, V. A. Miransky and I. A. Shovkovy, Nucl. Phys. B {\bf 462} (1996) 249-290.


\bibitem{KY} K. Kohri and S. Yamada, Phys. Rev. D {\bf 65} (2002) 043006. 








\bibitem{Hecht} E.~Hecht, ``{\it Optics, fourth edition}", Addison-Wesley (2001).
\bibitem{Fox} M.~Fox, ``{\it Optical properties of solids, second edition}", Oxford University Press (2010).

\bibitem{KMW} D. E. Kharzeev, L. D. McLerran, and H. J. Warringa, Nucl. Phys. A {\bf 803} (2008) 227-253. 
\bibitem{Sko} V. Skokov, A. Y. Illarionov and V. Toneev, Int. J. Mod. Phys. A {\bf 24} (2009) 5925-5932; 
A. Bzdak and V. Skokov, Phys. Lett. B {\bf 710} (2012) 171-174. 
\bibitem{DH} W.~T. Deng and X.~G. Huang, Phys. Rev. C {\bf 85} (2012) 044907.


\bibitem{Itakura_PIF} K. Itakura, ``{\it Strong Field Physics in 
High-Energy Heavy-Ion Collisions}", in 
``Proceedings of International 
Conference on Physics in Intense Fields (PIF2010)," 
(K. Itakura, et al. (eds.)) 24-26 November 2010, 
KEK, Tsukuba, Japan, available from 
{\tt http://ccdb5fs.kek.jp/tiff/2010/1025/1025013.pdf}


\bibitem{TD} C.~Thompson and R.~C.~Duncan, 
Mon.\ Not.\ Roy.\ Astron.\ Soc.\  {\bf 275} (1995) 255; 
{\it ibid.}, Astrophys.\ J.\  {\bf 473} (1996) 322.


\bibitem{MagnetarReview}A.~K.~Harding and D.~Lai, 
Rept.\ Prog.\ Phys.\  {\bf 69} (2006) 2631. 
{\tt arXiv:\,0606.674\,[astro-ph]}.

\bibitem{Adl70} S.~L.~Adler, et al., 
Phys. Rev. Lett. {\bf 25} (1970) 1061-1065. 
\bibitem{Bar} M.~G.~Baring, ``{\it Photon Splitting and Pair Conversion in Strong Magnetic Fields,}''  
AIP Conf.\ Proc.\  {\bf 1051} (2008) 53. {\tt arXiv:\,0804.0832\,[astro-ph]}.



\bibitem{Tu}K.~Tuchin,
Phys.\ Rev.\  C {\bf 82} (2010) 034904; 
{\it ibid.}, 
Phys.\ Rev.\ C {\bf 83}  (2011) 017901; 
{\it ibid.}, 
``{\it Electromagnetic radiation by quark-gluon plasma in magnetic field}", 
{\tt arXiv:\,1206.0485\,[hep-ph]}.

\bibitem{IHhbt} K. Itakura and K. Hattori, ``{\it Effects of extremely strong 
magnetic field on photon HBT interferometry}", PoS {\it WPCF2011} (2011) 042. 
{\tt arXiv:\,1206.3022\,[nucl-th]}.



\bibitem{TopicalReviewLASER}
Euro. Phys. J. D, Volume 55, Number 2 / November 2009 
``{\it Topical issue on Fundamental physics and ultra-high laser fields}".

\end{thebibliography}
